\documentclass[
reprint,
amsmath,
amssymb,
aps,
prb,
longbibliography,
noeprint
]{revtex4-2}
\usepackage{graphicx}
\usepackage{dcolumn}
\usepackage{bm}
\usepackage{needspace}

\usepackage{hyperref}
\usepackage{xcolor}
\hypersetup{
    colorlinks,
    linkcolor={red!50!black},
    citecolor={blue!50!black},
    urlcolor={blue!80!black}
}
\usepackage[capitalise]{cleveref}

\newenvironment{myfont}{\fontfamily{phv}\selectfont}{\par}
\makeatletter
\renewcommand\frontmatter@abstractwidth{\dimexpr\textwidth-0.5in\relax}
\makeatother

\newcommand{\NiXFe}{\ensuremath{ \mathrm{Ni}_x \mathrm{Fe}_{100-x} }}
\newcommand{\NiFe}[2]{\ensuremath{ \mathrm{Ni}_{#1} \mathrm{Fe}_{#2} }}

\newcommand{\dFM}{\ensuremath{ d_{\mathrm{FM}} }}

\newcommand{\Ms}{\ensuremath{ M_{\mathrm{s}} }}

\newcommand{\thetaAHE}{\ensuremath{ \vartheta_{\mathrm{AHC}} }}
\newcommand{\thetaSHE}{\ensuremath{ \vartheta_{\mathrm{SHE}} }}
\newcommand{\thetaAHT}{\ensuremath{ \vartheta_{\mathrm{AHT}} }}
\newcommand{\thetaSAHE}{\ensuremath{ \vartheta_{\mathrm{SAHE}} }}
\newcommand{\thetaX}{\ensuremath{ \vartheta_{\mathrm{AHS}} }}
\newcommand{\thetaSHT}{\ensuremath{ \vartheta_{\mathrm{SHT}} }}
\newcommand{\thetaPHT}{\ensuremath{ \vartheta_{\mathrm{PHT}} }}

\newcommand{\JFM}{\ensuremath{ J_{\mathrm{FM}} }}

\newcommand{\Idc}{\ensuremath{ I_{\mathrm{dc}} }}

\newcommand{\Ayz}{\ensuremath{ A_{\mathrm{yz}} }}

\begin{document}

\preprint{APS/123-QED}

\title{\myfont Self-generated spin-orbit torque driven by anomalous Hall current}

\author{\myfont Eric Arturo Montoya$^{1,2,\dagger,*}$}
\author{\myfont Xinyao Pei$^{1\dagger}$}
\author{\myfont Ilya N. Krivorotov$^{1,}$}

\email{Corresponding Authors. Emails: eric.montoya@utah.edu, ilya.krivorotov@uci.edu\\$\dagger$ These authors contributed equally.}

\affiliation{\myfont $^1$ Department of Physics and Astronomy, University of California, Irvine, California 92697, USA}
\affiliation{\myfont $^2$ Department of Physics and Astronomy, University of Utah, Salt Lake City, Utah 84112, USA}

\begin{myfont}

\begin{abstract}

\textbf{Spin-orbit torques enable energy-efficient manipulation of magnetization by electric current and hold promise for applications ranging from nonvolatile memory to neuromorphic computing. 
Here we report the discovery of a giant spin-orbit torque induced by anomalous Hall current in ferromagnetic conductors. 
This anomalous Hall torque is self-generated as it acts on magnetization of the ferromagnet that engenders the torque. 
The magnitude of the anomalous Hall torque is sufficiently large to fully negate magnetic damping of the ferromagnet, which allows us to implement a microwave spin torque nano-oscillator driven by this torque.
The peculiar angular symmetry of the anomalous Hall torque favors its use over the conventional spin Hall torque in coupled nano-oscillator arrays. 
The universal character of the anomalous Hall torque makes it an integral part of the description of coupled spin transport and magnetization dynamics in magnetic nanostructures.
}
\end{abstract}

\maketitle

\end{myfont}

Spin-orbit torques (SOTs) are relativistic quantum effects at the heart of spintronic devices such as nonvolatile magnetic memory \cite{liuSpinTorqueSwitching2012}, microwave nano-oscillators \cite{haidarSingleLayer2019,zahedinejadMemristiveControl2022,demidov_SHO_2012,liuMagneticOscillations2012,duan_nanowireSHO_2014}, ultrafast spectrum analyzers \cite{litvinenko_ultrafast_2022} and terahertz emitters \cite{seifert_efficientemiiter_2016}. 
SOTs generated by electric charge currents via spin-orbit coupling (SOC)  act on magnetic order parameters of ferromagnets (FM) \cite{ando_electric_manipulation_2008, huang_novel_2021} and antiferromagnets (AFM) \cite{wadley_AFMswitching_2016, yan_SOT_AFM_2022, higo_perpendicular_2022}. 
Practical applications of SOTs are based on their ability to efficiently drive nonlinear magnetic dynamics such as switching \cite{mironPerpendicularSwitching2011, petrovic_quantum_spintorque_2021}, auto-oscillations \cite{liuMagneticOscillations2012, demidov_SHO_2012,duan_nanowireSHO_2014, hache_bipolar_2020, hamadeh_simultaneous_2023} and skyrmion transport \cite{woo_SOT_skyrmionDynamics_2016}. 

SOTs are usually studied in multilayers of FM and non-magnetic materials (NM). 
The most explored SOTs are spin Hall torque (SHT) originating from SOC in NM layers \cite{ando_electric_manipulation_2008,liuSpinTorqueSwitching2012, liuMagneticOscillations2012, li_charge-spin_2021-1, ikebuchi_crystal_2022} and Rashba torque (RT) induced by SOC at FM$|$NM interfaces \cite{mironPerpendicularSwitching2011}.  
Recently, several other types of SOTs emerged, including planar Hall torque (PHT) \cite{safranskiPlanarHall2020}, spin-swapping torque \cite{ pauyac_SpinSwapping_2018} and interface-generated torques \cite{belashchenko_interfacial_2020, hibino_giant_2021-1}.
Other types of torques such as orbital torque \cite{lee_OrbitalTorque_2021} and spin Seebeck torque \cite{safranski_SpinCaloritronicSNO_2017} can be present in FM$|$NM heterostructures as well. 
The multiplicity of coexistent spin torques of similar magnitude but different origin and symmetry \cite{garello_symmetry_2013, dc_observation_2023,gibbons_large_2022} 
is a challenge for the development
of spintronic devices.

SOTs universally present in all magnetic systems are especially important for the formulation of a comprehensive theory of current-driven magneto-dynamics.  
Here we report the discovery of a universal SOT driven by spin and charge currents of anomalous Hall origin in FM conductors. 
This anomalous Hall torque (AHT) is self-generated because it acts on magnetization of the FM producing the torque. 
The magnitude of AHT is sufficiently large to cancel magnetic damping of the FM and thereby drive persistent auto-oscillations of its magnetization. 
AHT exhibits an unusual angular symmetry that favors its applications in coupled spin torque nano-oscillators (STNOs) for neuromorphic signal processing \cite{ romera_binding_2022}.

\bigskip
\begin{figure*}[tbph]
\centering
 \includegraphics[width= 1.0\textwidth]{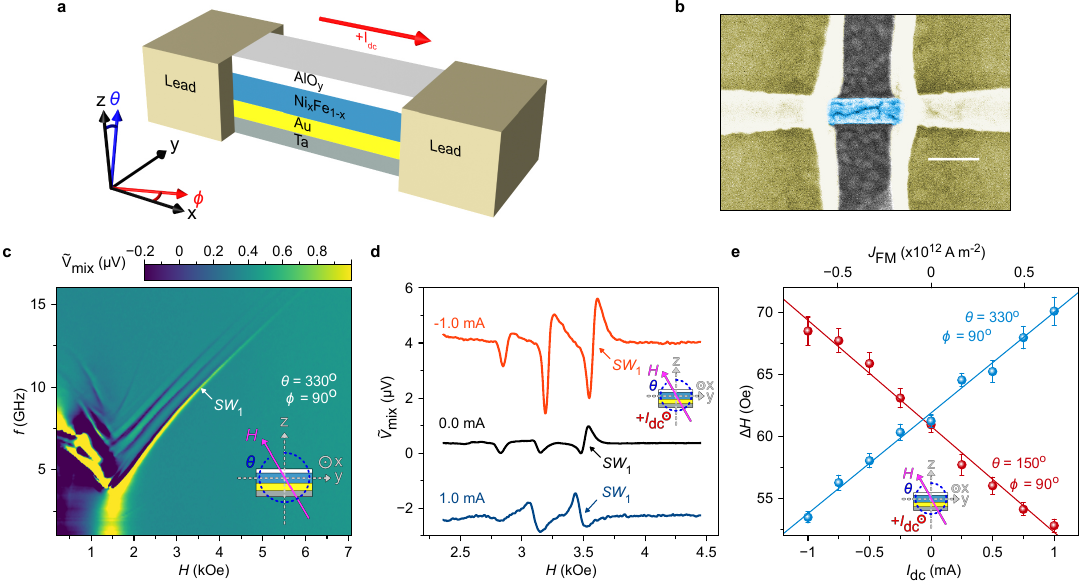}%
 \caption{\textbf{Device geometry and spin-torque ferromagnetic resonance.} \textbf{(a)} Schematic of the nanowire device with Cartesian ($x,y,z$) and spherical ($\theta$, $\phi$) coordinate systems. 
 \textbf{(b)} Scanning electron micrograph of the nanowire device with 100 nm scale bar. The blue and yellow colors indicate the nanowire active region and electric leads, respectively.  Etching-induced roughness of negative e-beam resist on top of the wire is seen in the active region (Methods).
 \textbf{(c)} ST-FMR signal $\tilde{V}_\mathrm{mix}$ measured as a function of the drive frequency $f$ and magnetic field $H$ applied in the $yz$-plane for the $\mathrm{Ta(3\,nm)}\allowbreak|\mathrm{Au(3\,nm)}\allowbreak| \NiFe{70}{30}\mathrm{(5\,nm)}\allowbreak|\mathrm{AlO_y(2\,nm)}$ nanowire device at $I_\mathrm{dc}=0$. 
 The range of $\tilde{V}_\mathrm{mix}$ displayed is clipped compared to the full range of $-1.0$ to $19\,\mathrm{\mu V}$ in order to clearly show the higher order spin wave modes. 
 \textbf{(d)} ST-FMR spectra $\tilde{V}_\mathrm{mix}(H)$ measured at $f=10$\,GHz and $H$ applied in the $yz$-plane ($\theta=330^{\circ}$ and $\phi=90^{\circ}$) for three values of $I_\mathrm{dc}$. 
 \textbf{(e)} Linewidth $\Delta H$ of the lowest-frequency spin wave mode $SW_\mathrm{1}$ versus $\Idc$ for two directions of $H$ in the $yz$-plane.
\label{fig:Schematics_ST-FMR}}%
 \end{figure*}

\begin{large}
\noindent \textbf{\myfont Anomalous Hall torque}
\end{large}

We measure SOTs in substrate$\allowbreak||\mathrm{Ta(3\,nm)}\allowbreak|\mathrm{Au(3\,nm)}\allowbreak| \NiXFe\mathrm{(5\,nm)}\allowbreak|\mathrm{AlO_y(2\,nm)}$ nanowire devices (Methods and Supplementary Note 1) schematically shown in \cref{fig:Schematics_ST-FMR}(a) along with the coordinate system.
\Cref{fig:Schematics_ST-FMR}(b) shows a scanning electron micrograph of such a device. The 171\,nm gap between two $\mathrm{Ta}(5\,\mathrm{nm})| \allowbreak \mathrm{Au}(40\,\mathrm{nm})| \allowbreak \mathrm{Ta}(15\,\mathrm{nm})$ leads attached to a 56\,nm wide, 30\,$\mu$m long nanowire defines the active region where high electric current density is applied.
The $\NiXFe$ FM layer allows us to tune the magnitude and sign of the anomalous Hall effect (AHE) via variation of the Ni content $x$.
The NM under-layer NM$_1(=\mathrm{Ta}|\mathrm{Au})$ acts as a good sink of spin current generated in the FM and impingent upon the NM$_1$ layer. 
The NM$_2(=\mathrm{AlO_\mathrm{y}}$) capping layer is a poor sink of spin current generated in the FM (Supplementary Note 2). 
The different spin current absorption properties at the top and bottom surfaces of the FM film are essential for the generation of AHT. 

We use spin torque ferromagnetic resonance (ST-FMR) \cite{goncalvesSpinTorque2013} to measure damping-like SOTs acting on magnetization $\mathbf{M}=\Ms \mathbf{m}$ of the FM layer, where  $\mathbf{m}=(m_x, m_y, m_z)$ is a unit vector and $\Ms$ is the FM saturation magnetization (Methods). 
Room-temperature ST-FMR spectra of the rectified voltage $\tilde{V}_\mathrm{mix}$ versus external magnetic field $H$ and frequency $f$ of the applied microwave current reveal multiple spin wave resonances shown in \cref{fig:Schematics_ST-FMR}(c) \cite{safranskiSpinOrbit2019}.
The discrete spectrum of spin wave eigenmodes arises from geometric confinement to the active region \cite{duan_nanowireSHO_2014}.
The data in \cref{fig:Schematics_ST-FMR}(c) reveal that $\mathbf{M}$ is saturated for $H>1.5$\,kOe \cite{duan_spin-wave_2014}.

A damping-like SOT from the direct current $\Idc$ applied to the device tunes the effective magnetic damping $\alpha(\Idc)$ of the FM. 
Since the linewidth $\Delta H$ of a spin wave resonance is linear in $\alpha$, measurements of $\Delta H(\Idc)$ probe $\alpha(\Idc)$.
\Cref{fig:Schematics_ST-FMR}(d) shows that $\Idc$ modifies $\Delta H$ for $H$ applied in the $yz$-plane ($\theta=330^{\circ}$, $\phi=90^{\circ}$), revealing a damping-like SOT.
In these measurements, $H$ exceeds the saturation field, which forces $\mathbf{M}$ to be nearly parallel to the field. 
\Cref{fig:Schematics_ST-FMR}(e) shows that $\Delta H$ of the lowest-frequency mode ($SW_\mathrm{1}$) is linear in $\Idc$, confirming a damping-like SOT in the $yz$-plane with efficiency characterized by the slope $d \Delta H / d \Idc$. 

\begin{figure*}[tbph]
\centering
 \includegraphics[width= 1.0\textwidth]{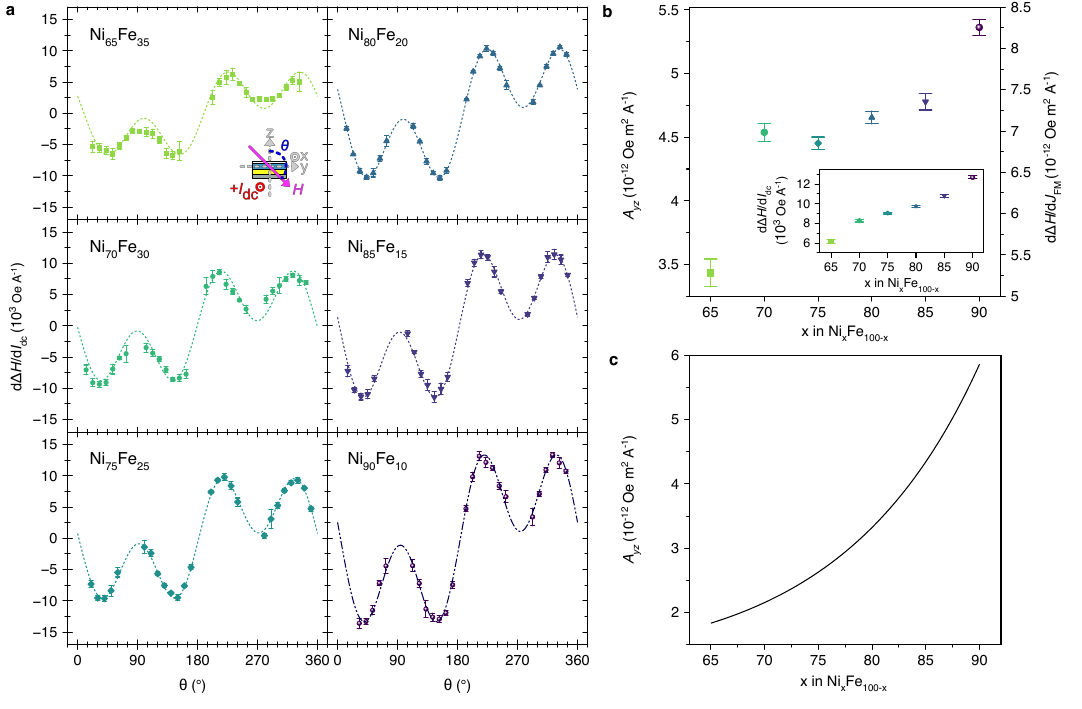}%
 \caption{\textbf{Angular and alloy composition dependence of anomalous Hall torque.} 
 \textbf{(a)} Dependence of the AHT efficiency $d \Delta H / d \Idc$ on the direction $\theta$ of magnetization in the $yz$-plane for six $\NiXFe$ alloy compositions ($\phi = 90^\circ$). 
 \textbf{(b)} Dependence of the maximum AHT efficiency $d \Delta H / d \JFM$ at $\theta=325^\circ$ and $A_{yz}$ on the alloy composition $x$.
 The inset shows $d \Delta H / d \Idc$ at $\theta=325^\circ$ versus $x$. 
 \textbf{(c)} Theoretical dependence of $A_{yz}$ on the alloy composition.
\label{fig:AHT}}%
 \end{figure*}

\Cref{fig:AHT}(a) shows the dependence of $d \Delta H / d \Idc$ on the direction of $\mathbf{M}$ in the $yz$-plane for six $\NiXFe \allowbreak$ alloy compositions.  
The expected angular dependence of the SHT- and RT-induced damping in the $yz$-plane is $\sin(\theta)$ while the measured angular dependence is $\sin(\theta)\cos^2(\theta)=  \frac{1}{4}[\sin(\theta)+\sin(3\theta)]$.  
The lines in \cref{fig:AHT}(a) are best fits to $A[\sin(\theta)+\sin(3\theta)] +B\sin(\theta)$ with fitting parameters $A$ and $B$, where $B$ is constrained to be identical for all alloy compositions $x$. 
The fitting gives $B \ll A$ for all $x$, consistent with small spin Hall angle in Au \cite{safranskiSpinOrbit2019}.

The observed angular dependence in \cref{fig:AHT}(a) coincides with that recently predicted for self-generated AHT
\cite{ochoaSelfinducedSpinorbit2021a}. 
This torque is predicted to arise from $z$-component of anomalous Hall spin and charge currents in a FM conductor, and is applied to magnetization of the same FM conductor. 
The theory predicts AHT-induced angular-dependent linewidth \cite{ochoaSelfinducedSpinorbit2021a}:
\begin{equation}\label{eq:OchoaAngle}
    -d \Delta H / d \JFM = A_{yz}  \left[ \sin (\theta) + \sin (3 \theta) \right] \sin (\phi),
\end{equation}
where $\JFM$ is the $x$-component of the electric charge current density 
$\bf{\JFM}$
in the FM, and $A_{yz}$ characterizes the AHT-induced damping efficiency.
\Cref{eq:OchoaAngle} is not a truncated expansion in odd harmonics because both $\sin(\theta)$ and $\sin(3\theta)$ terms have identical amplitudes.
The unusual angular dependence of \cref{eq:OchoaAngle} describes our experimental data in \cref{fig:AHT}(a) well, suggesting that AHT is the dominant SOT in our system.

Using \cref{fig:AHT}(a), we plot the maximum of $d \Delta H / d \Idc$ (observed at $\theta_{\mathrm{max}} = 360^\circ - \sin^{-1}(1/\sqrt(3) \approx 325^\circ)$ versus $x$ in the inset of \cref{fig:AHT}(b). 
Measurements of electrical conductivity of the individual FM and NM layers (Supplementary Note 3) allow us to calculate $\JFM$ and plot $d \Delta H / d \JFM$ at $\theta=325^\circ$ versus $x$ in \cref{fig:AHT}(b). 
It is clear from \cref{eq:OchoaAngle} that $A_{yz}$ is proportional to $d \Delta H / d \JFM$ at $\theta_{\mathrm{max}}$: $A_{yz} = \frac{3\sqrt{3}}{8}\, d \Delta H / d \JFM$. 
The data in \cref{fig:AHT}(b) reveal that $A_{yz}$ increases with increasing concentration of Ni in the alloy.
Supplementary Note 4 shows that the measured AHT magnitude is frequency- and field-independent consistent with the AHT theory \cite{ochoaSelfinducedSpinorbit2021a}.
 
\bigskip
\begin{large}
\noindent \textbf{\myfont Anomalous Hall torque nano-oscillator} 
\end{large}

The damping-like AHT is sufficiently large to cancel the intrinsic damping of $\NiXFe$ and thereby induce persistent auto-oscillations of $\mathbf{M}$ \cite{ slavin_nonlinearSTO_2009}. 
To demonstrate a STNO driven by AHT, we make a 56\,nm-wide nanowire device with a 297\,nm long active region from a substrate$|| 
\allowbreak \mathrm{Ta}(5\, \mathrm{nm})| \allowbreak \NiFe{90}{10} (5\, \mathrm{nm})| \allowbreak \mathrm{AlO_{y}}(2\, \mathrm{nm})$ multilayer.
Compared to the device in \cref{fig:Schematics_ST-FMR}(b), we remove the Au layer in order to (i) increase $\JFM$ and thus the magnitude of AHT and (ii) increase anisotropic magnetoresistance (AMR, Supplementary Note 5) and thus the output microwave power.
Resistivity measurements of $\mathrm{Ta}(5\, \mathrm{nm})$ and $\NiFe{90}{10} (5\, \mathrm{nm})$ layers (Supplementary Note 3) show that 87$\%$ of $\Idc$ flows in the $\NiFe{90}{10} (5\, \mathrm{nm})$ layer. 

\begin{figure*}[htbp]
\centering
 \includegraphics[width= 1.0\textwidth]{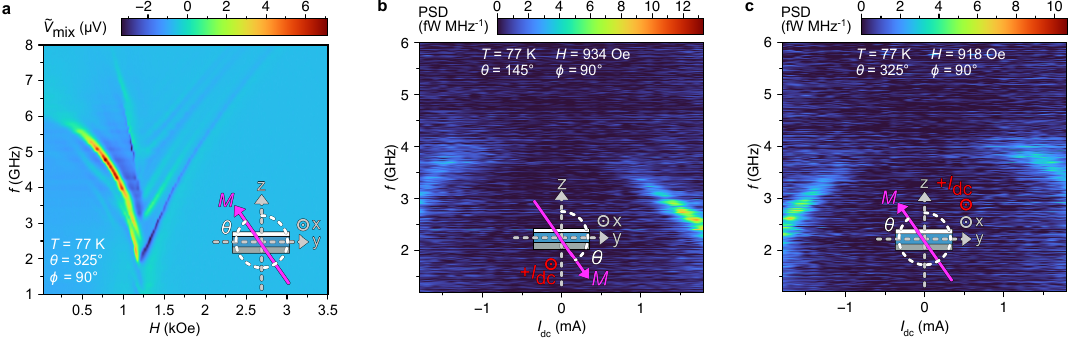}%
 \caption{\textbf{Anomalous Hall torque nano-oscillator.} \textbf{(a)} ST-FMR signal $\tilde{V}_\mathrm{mix}$ measured at $\Idc=0$ as a function of the drive frequency $f$ and bias magnetic field $H$ applied in the $yz$-plane ($\theta=325^{\circ}$ and $\phi=90^{\circ}$) for the $\mathrm{Ta}(5\, \mathrm{nm})| \NiFe{90}{10} (5\, \mathrm{nm})|\mathrm{AlO_{y}}(2\, \mathrm{nm})$ nanowire STNO. \textbf{(b,\,c)} Power spectral density (PSD) of the microwave signal generated by the STNO as a function of frequency $f$ and $\Idc$  measured at \textbf{(b)} $\theta=145^{\circ}$, $\phi=90^{\circ}$ and \textbf{(c)} $\theta=325^{\circ}$, $\phi=90^{\circ}$.
\label{fig:Emission}}%
 \end{figure*}

Spin pumping measurements show that the Ta layer acts as an efficient spin sink in agreement with previous studies (Supplementary Note 2). 
Therefore, the spin current absorption asymmetry at the top and bottom FM surfaces necessary for the generation of AHT is realized \cite{ochoaSelfinducedSpinorbit2021a}.
\Cref{fig:Emission}(a) shows ST-FMR spectra for this device measured for $H$ applied in the $yz$-plane at $\theta = 325^\circ$ and $\phi = 90^\circ$. 
The spectra show multiple spin wave eigenmodes and reveal that $\mathbf{M}$ is saturated for $H>1.25 \, \mathrm{kOe}$. 

\Cref{fig:Emission}(b) shows the spectra of microwave signal spontaneously emitted by the STNO in response to $\Idc$ (Methods). 
To maximize the AHT-induced antidamping, these measurements are made for $H = 934 \, \mathrm{Oe}$ applied in the $yz$-plane at $\theta = 145^\circ$ and $\phi = 90^\circ$. 
The value of $H$ below the saturation field creates a small deviation of the equilibrium direction of $\mathbf{M}$ from the $yz$-plane. 
Such a deviation increases the conversion efficiency of magnetization auto-oscillations into AMR oscillations and boosts the output power \cite{duan_nanowireSHO_2014}.
To mitigate detrimental effects of Ohmic heating on the STNO operation at large values of $\Idc$  \cite{demidov_fluctuation_control_2011}, the data in \cref{fig:Emission} are collected at $T=$\,77 K. 

\Cref{fig:Emission}(b) reveals spontaneous microwave signal generation by the STNO at the lowest-frequency spin wave mode for $\Idc$ exceeding the critical value of approximately 0.75\,mA. 
This auto-oscillatory mode shows red frequency shift with increasing $\Idc$ primarily due to Ohmic heating that decreases $M_s$ and magnetic shape anisotropy. 
For $\Idc<0$, we observe magnoise signal arising from incoherent thermal magnons induced by Ohmic heating \cite{demidov_fluctuation_control_2011}. 
The amplitude of  magnoise is significantly lower than that due to the coherent auto-oscillations at $\Idc>0$. 
\Cref{fig:Emission}(c) shows that reversal of $H$ leads to auto-oscillatory dynamics for negative $\Idc$, as expected from \cref{eq:OchoaAngle}.

Our observation of large AHT in the devices with different NM$_1$ layers $(\mathrm{Ta}|\mathrm{Au}$ and $\mathrm{Ta})$ is consistent with the mechanism of AHT, which only requires asymmetry of spin current absorption at the top and bottom surfaces of the FM film \cite{ochoaSelfinducedSpinorbit2021a} as discussed in the next section.

\bigskip
\begin{large}
\noindent \textbf{\myfont Discussion}
\end{large}

\begin{figure*}[t!bhp]
\centering
 \includegraphics[width= 1.0\textwidth]{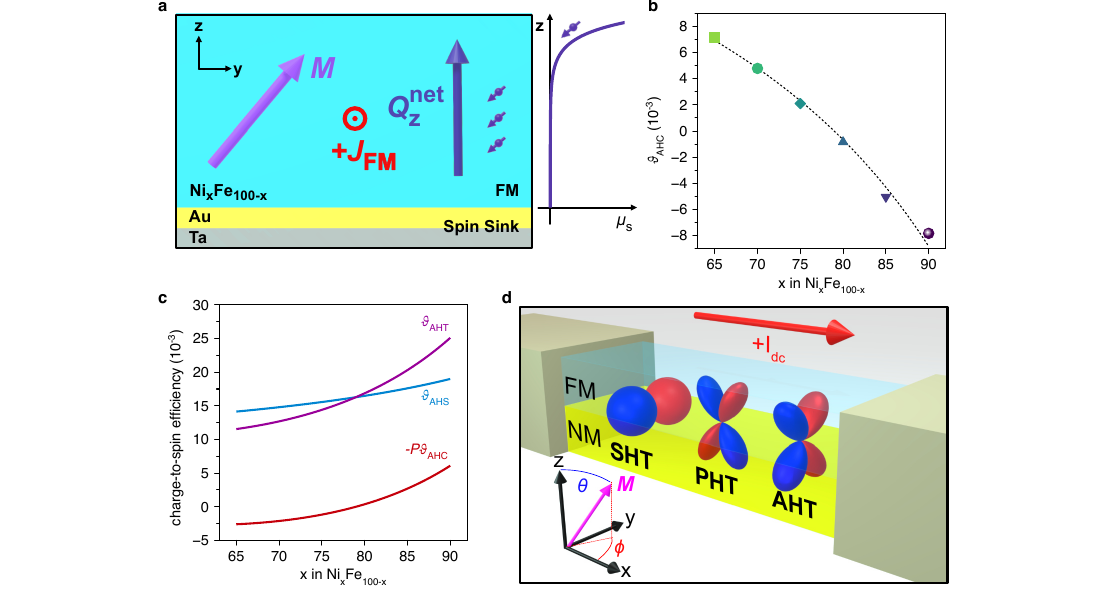}%
 \caption{\textbf{Anatomy of anomalous Hall torque.} \textbf{(a)} Schematic showing the flow of out-of-plane spin current density $Q_\mathrm{z}^\mathrm{net}$ driven by in-plane charge current density $\JFM$, and the resultant spin accumulation $\mu_s(z)$ in the FM layer for $\JFM>0$. 
 \textbf{(b)} Anomalous Hall angle $\thetaAHE$ versus $x$ in $\NiXFe$. 
 \textbf{(c)} Anomalous Hall torque efficiency $\thetaAHT$ (purple) and contributions to $\thetaAHT$ from anomalous Hall spin current $\thetaX$ (blue) and the counter-flow current $-P \thetaAHE$ (red). \textbf{(d)} Schematic of the angular dependence of damping-like SOTs in a NM$|$FM system: SHT, PHT and AHT. Here red (blue) corresponds to negative (positive) current-induced damping for $\Idc>0$ and positive SOT efficiency parameters $\thetaSHT$, $\thetaPHT$ and $\thetaAHT$.
\label{fig:Anatomy}}%
\end{figure*}

The origin of the self-generated AHT \cite{ochoaSelfinducedSpinorbit2021a} is schematically illustrated in \cref{fig:Anatomy}(a). 
A charge current density $\JFM$ in the FM generates a transverse anomalous Hall spin current density $\mathbf{Q}_\mathrm{z}$ flowing along the $z$-axis \cite{aminIntrinsicSpin2019}. 
The origin of $\mathbf{Q}_\mathrm{z}$ is SOC in the FM, its magnitude $Q_\mathrm{z} = |\mathbf{Q}_\mathrm{z}|$ depends on the FM electronic band structure and its direction $\mathbf{Q}_\mathrm{z}/Q_\mathrm{z}$ is collinear with the spin angular momentum it carries.
The $\mathbf{Q}_\mathrm{z}$ vector generally has components parallel and perpendicular to $\mathbf{M}$ \cite{aminIntrinsicSpin2019}. 
Since only the $\mathbf{Q}_\mathrm{z}$ component parallel to $\mathbf{M}$ contributes to the damping-like AHT \cite{ochoaSelfinducedSpinorbit2021a}, we neglect the perpendicular component and assume $\mathbf{Q}_\mathrm{z} || \mathbf{M}$. 
 
The magnitude of $\mathbf{Q}_\mathrm{z}$ depends on the direction of $\mathbf{M}$ \cite{aminIntrinsicSpin2019}:
$Q_\mathrm{z}=- \frac{\hbar}{2 |e|} \thetaX m_y \JFM$
where $\hbar$ is reduced Planck constant, $e$ is electron charge and $\thetaX$ is a material-specific dimensionless parameter  describing the conversion efficiency of $\JFM$ into $Q_\mathrm{z}$.
Here $\thetaX$ is a spin-sector analog of the anomalous Hall angle $\thetaAHE$ in the charge sector, 
where $\thetaAHE$ quantifies the conversion of $\JFM$ into transverse anomalous Hall charge current density $J_\mathrm{z}=\thetaAHE m_y \JFM$. In the literature \cite{aminIntrinsicSpin2019}, $\thetaX$ is usually split onto two terms: spin Hall angle $\thetaSHE$ and spin anomalous Hall angle $\thetaSAHE$. 
For the physics of AHT, only the algebraic sum $\thetaX=\thetaSHE+\thetaSAHE$ is relevant \cite{aminIntrinsicSpin2019, ochoaSelfinducedSpinorbit2021a}.

While $\thetaX$ characterizes the anomalous Hall spin current in an infinite FM, another contribution to the transverse spin current is present in a FM film \cite{taniguchiSpinTransferTorques2015,ochoaSelfinducedSpinorbit2021a}  due to $J_\mathrm{z}$ generating charge accumulation at the top and bottom surfaces of the FM film (Supplementary Note 6). 
This charge accumulation drives a counter-flow charge current density $J_\mathrm{z}^\mathrm{cf}=-J_\mathrm{z}$ along the $z$-axis. 
The counter-flow charge current is spin-polarized along $\mathbf{M}$, which gives rise to a counter-flow spin current density $Q_\mathrm{z}^\mathrm{cf} = \frac{\hbar}{2 |e|} P \thetaAHE m_y \JFM$ \cite{aminIntrinsicSpin2019}, where $-1\leq P\leq1$ is intrinsic spin polarization of charge current in the FM.

The magnitude of the net transverse spin current density in the FM film $Q_\mathrm{z}^\mathrm{net}$ is the sum of $Q_\mathrm{z}$ and $Q_\mathrm{z}^\mathrm{cf}$:
\begin{equation}\label{eq:Qznet}
Q_\mathrm{z}^\mathrm{net}= - \frac{\hbar}{2 |e|} (\thetaX -  P \thetaAHE) \sin{(\theta)}\sin{(\phi)} \JFM.
\end{equation}
Asymmetric spin current sinks at the top and bottom surfaces of the FM film give rise to an asymmetric spatial profile of spin accumulation $\boldsymbol{\mu_s}(z)$ driven by $\mathbf{Q}_\mathrm{z}^\mathrm{net}$, as illustrated in \cref{fig:Anatomy}(a).
A significant spin accumulation $\boldsymbol{\mu_s} || \mathbf{Q}_\mathrm{z}^\mathrm{net}$ is created at the FM interface with a poor spin sink NM$_2$, and $\boldsymbol{\mu_s} \approx 0$ at the interface with a good spin sink NM$_1$. 
This creates a net current-driven spin accumulation within the FM film $\boldsymbol{\bar{\mu}_s} \sim \mathbf{Q}_\mathrm{z}^\mathrm{net}$.

For the direction of $\bf{\JFM}$ shown in \cref{fig:Anatomy}(a), the magnetic moment of $\boldsymbol{\bar{\mu}_s}$ is opposite to $\mathbf{M}$. 
Such spin accumulation can give rise to antidamping spin torque applied to $\mathbf{M}$. 
The mechanism of such torque can be either  magnonic or induced by a symmetry-breaking SOC at the FM$|$NM$_2$ interface \cite{ochoaSelfinducedSpinorbit2021a}. 
In the magnonic mechanism, $\boldsymbol{\bar{\mu}_s}$ relaxes via generation of an non-equilibrium magnon cloud, which in turn interacts with the quasi-uniform dynamic mode of $\mathbf{M}$ and thereby decreases the mode's effective damping  \cite{ochoaSelfinducedSpinorbit2021a, safranski_SpinCaloritronicSNO_2017}. 
The conversion efficiency of  $\boldsymbol{\bar{\mu}_s}$ into the damping-like AHT depends on the direction of $\mathbf{M}$ and the symmetry of the magnetic system \cite{ochoaSelfinducedSpinorbit2021a}. 
For a thin-film FM with rotation symmetry around $z$-axis, the conversion efficiency of $\boldsymbol{\bar{\mu}_s}$ into current-induced damping $d \Delta H / d \JFM$ is proportional to $m_z^2=\cos^2(\theta)$ \cite{ochoaSelfinducedSpinorbit2021a}.
Assuming maximum efficiency of spin angular momentum transfer from $\boldsymbol{\bar{\mu}_s}$ to the quasi-uniform mode of $\mathbf{M}$ allowed by symmetry (Supplementary Note 7):
\begin{equation}\label{eq:ConvEff}
 d \Delta H / d \JFM = - \frac{\hbar\thetaAHT}{2 |e| M_\mathrm{s} \dFM}  \cos^2(\theta) \sin(\theta) \sin(\phi),
\end{equation}
where $\dFM$ is the FM layer thickness and $\thetaAHT$ is the dimensionless AHT coefficient: 
\begin{equation}\label{eq:Ochoa}
    \thetaAHT = \thetaX - P \thetaAHE.
\end{equation}
The derivation of \cref{eq:ConvEff} neglects the effect of magnetic anisotropy on the linewidth $\Delta H$. 
This is a reasonable approximation because $H$ used in the measurements of $\Delta H$ significantly exceeds the saturation field.

The angular dependence of the current-induced damping given by \cref{eq:ConvEff} is in full agreement with our experimental data in \cref{fig:AHT}(a) described by \cref{eq:OchoaAngle}. 
\Cref{eq:Qznet} clearly shows that the observed damping-like torque is driven by a combination of the anomalous Hall spin and charge currents in the FM film, which justifies naming this torque the self-generated AHT.

Comparison of \cref{eq:OchoaAngle} and \cref{eq:ConvEff} gives the AHT-induced damping efficiency (Supplementary Note 7):
\begin{equation}\label{eq:Ayz}
    \Ayz = \frac{\hbar}{8 \left| e \right|} \frac{\thetaAHT}{\Ms \dFM}.
\end{equation}
The current-induced damping described by \cref{eq:ConvEff} corresponds to the AHT vector $\mathbf{T}_\mathrm{AHT}$ applied to $\mathbf{M}$ \cite{ochoaSelfinducedSpinorbit2021a}:
\begin{equation}\label{eq:TorqeForm}
\mathbf{T}_\mathrm{AHT}= -\gamma
\frac{\hbar}{2 |e|}
\frac{\thetaAHT}{\dFM} m_z (\bf{m} \times \hat{z} \times \bf{m})(\bf{m} \cdot (\hat{z} \times \bf{\JFM})),
\end{equation}
where $\gamma$ is the absolute value of the electron gyromagnetic ratio.
This vector form of AHT can be directly used in micromagnetic simulations in the presence of AHT (Supplementary Note 8).

The dependence of $\Ayz$ on the FM alloy composition $x$ can be calculated from \cref{eq:Ochoa} and \cref{eq:Ayz}, provided the dependence of $\thetaX$, $\thetaAHE$, $P$, and $\Ms$ on $x$ is known. 
We measure $\thetaAHE(x)$  for our samples as shown in \cref{fig:Anatomy}(b) 
 (Supplementary Note 5). 
The function $M_\mathrm{s}(x)$ and the values of $\thetaX$ and $P$ for Fe and Ni are known \cite{aminIntrinsicSpin2019}, which allows us to approximate $\thetaX(x)$ and $-P(x)\thetaAHE(x)$ shown in \cref{fig:Anatomy}(c) along with their sum, $\thetaAHT(x)$, by interpolating functions (Supplementary Note 7).
This figure reveals that $\thetaAHT$ for $\NiXFe$ is predominantly determined by the anomalous Hall spin current $\sim\thetaX$. 
The counter-flow current contribution $-P\thetaAHE$ is smaller but non-negligible, and $\thetaAHT$ does not change sign as a function of $x$ in agreement with the data in \cref{fig:AHT}(b). 

\Cref{fig:AHT}(c) shows $\Ayz(x)$ calculated from \cref{eq:Ayz}. 
The agreement between the measured $\Ayz$ in \cref{fig:AHT}(b) and the predicted $\Ayz(x)$ in \cref{fig:AHT}(c)  is very good given that no fitting parameters are used in the calculation. 
The theory accurately captures the sign of $\Ayz$, is overall magnitude and its monotonic increase with $x$.  
Since the theory of \cref{eq:ConvEff} through \cref{eq:Ayz} captures the angular dependence, the magnitude, the sign and the alloy composition dependence of the measured damping-like SOT, we conclude that this SOT is indeed the self-generated AHT \cite{ochoaSelfinducedSpinorbit2021a}.  
The residual differences between the measured and predicted $\Ayz(x)$ are likely a result of the approximations used in the calculation of $\Ayz(x)$: (i) the use of interpolating functions for $\thetaX(x)$ and $P(x)$, (ii) neglecting magnetic anisotropy contributions to $\Delta H$ in \cref{eq:ConvEff} and (iii) approximating the top and bottom FM$|$NM interfaces by perfectly good and poor spin sinks, respectively. The critical current of the AHT-based STNO predicted by the theory (Supplementary Note 9) is similar to its measured value in \cref{fig:Emission}(b) as well.

\Cref{fig:Anatomy}(c) demonstrates that $\thetaX$ and $-P\thetaAHE$ can add constructively to enhance $\thetaAHT$. 
This holds promise for  engineering of magnetic materials with giant AHT, because materials with $\thetaX$ \cite{cespedes-berrocalCurrentInducedSpin2021} and $\thetaAHE$ \cite{liuGiantAnomalous2018} far exceeding those of $\NiXFe$ have recently been found. 
A systematic search aiming to maximize $\thetaAHT$ is likely to yield materials with giant AHT, which is very promising because the magnitude of AHT in $\NiXFe$ is already similar to that of the giant SHT in Pt \cite{ando_electric_manipulation_2008}.  

The mechanism of AHT is analogous to the mechanism of the self-generated PHT in FM conductors \cite{safranskiPlanarHall2020, ochoaSelfinducedSpinorbit2021a}. 
In the case of PHT, a transverse spin current in a FM along $z$-axis is generated by the SOC that gives rise to AMR and planar Hall effect (PHE). 
The dependence of this transverse planar Hall spin current on $\mathbf{m}$ is different from that of the anomalous Hall spin current given by \cref{eq:Qznet}, resulting in the angular dependence of PHT-induced damping different from that of \cref{eq:ConvEff} \cite{safranskiPlanarHall2020, ochoaSelfinducedSpinorbit2021a}.

\Cref{fig:Anatomy}(d) illustrates the dependence of current-induced damping on the direction of $\mathbf{M}$ due to SHT, PHT \cite{safranskiPlanarHall2020} and AHT (\cref{eq:ConvEff}) in a FM$|$NM bilayer. 
While the effect of SHT on damping is maximized for $\mathbf{M}$ along the $y$-axis, both AHT and PHT do not modify the damping for $\mathbf{M}$ in the $xy$-plane and along the $z$-axis. 
AHT is active in the $yz$-plane while PHT is active in the $xz$-plane. 
Given that AMR and AHE are generally non-zero in a FM conductor, we expect AHT and PHT to be universally present in FM systems with spin sink asymmetry. 
Therefore, AHT and PHT should be a part of the general description of SOTs in systems with FM conductors. Furthermore, AFM conductors have been shown to exhibit AMR and AHE \cite{nakatsuji_AHEinAFM_2015}; therefore, we expect PHT and AHT to be relevant for AFM spintronics.

The peculiar angular dependence of AHT and PHT makes these torques ideal candidates for STNO arrays coupled via spin waves \cite{awad_SNO_sync_2017}. 
Indeed, STNO coupling via spin waves demands a large out-of-plane component of $\mathbf{M}$ \cite{slavin_nonlinearSTO_2009, awad_SNO_sync_2017}. 
However, \cref{fig:Anatomy}(d) shows that SHT efficiency decreases when $\mathbf{M}$ rotates out of the sample plane, leading to large currents and reduced energy efficiency in SHT-based coupled STNO arrays. 
In contrast, STNOs based on AHT or PHT are expected to show maximum efficiency for $\mathbf{M}$ having a large out-of-plane component.
Coupled STNOs hold promise for neuromorphic signal processing \cite{ hassanLowBarrierMagnet2019, zahedinejadMemristiveControl2022}. 

The self-generated AHT described here is fundamentally different from previously studied torques induced by anomalous Hall currents. 
Anomalous Hall spin currents can be used to transfer angular momentum from one FM conductor to another via a NM spacer in a FM$_1|$NM$|$FM$_2$ multilayer, which gives rise to spin transfer torque induced by anomalous Hall currents \cite{taniguchiSpinTransferTorques2015, gibbonsReorientableSpin2018, bose_observation_2018, sunFieldFreeSwitching2019}.
Anomalous Hall current in a single FM conductor also gives rise to anomalous spin-orbit torque (ASOT) \cite{wangAnomalousSpin2019}. ASOT is equal in magnitude but opposite in direction at the top and bottom surfaces of the FM film, leading to fanning of $\mathbf{M}$ across the FM film thickness rather than to modification of the FM damping. 

Self-generated SOTs of origins different from AHT have been observed in several systems \cite{kurebayashiAntidampingSpin2014, safranskiSpinOrbit2019, wangAnomalousSpin2019,  zhuObservationStrong2020, liuElectricalSwitching2020, zhengFieldfreeSpinorbit2021, cespedes-berrocalCurrentInducedSpin2021} (Supplementary Note 8). 
A combination of Dresselhaus and Rashba SOC terms gives rise to self-generated SOTs in a (Ga,Mn)As FM semiconductor with broken bulk crystal inversion symmetry \cite{kurebayashiAntidampingSpin2014}. 
Field-free switching due to a self-generated SOT has been demonstrated in FM layers where a chemical composition gradient along the film thickness leads to bulk inversion symmetry breaking \cite{liuElectricalSwitching2020, zhengFieldfreeSpinorbit2021}.  
These torques have angular symmetry different from  AHT and, in contrast to AHT, modify the FM damping for $\mathbf{M}$ in the $xy$-plane \cite{cespedes-berrocalCurrentInducedSpin2021}.

\bigskip
\begin{large}
\noindent \textbf{\myfont Conclusions}
\end{large}

The universal AHT reported here opens new venues of research in spintronics and magnonics. 
The peculiar angular symmetry of AHT can drive unusual dynamics of magnetic textures such as skyrmions, vortices, and domain walls. 
Magnetization reversal process in non-volatile SOT memories is generally affected by AHT, and development of energy-efficient switching strategies that take advantage of AHT is needed. 
The symmetry of AHT is ideal for energy-efficient coupling of STNOs, and arrays of AHT-based STNOs can be explored for neuromorphic computing applications. 
A systematic search of materials supporting giant anomalous Hall spin and charge currents is likely to yield systems with very large AHT. 

AHT, along with SHT and PHT, forms a triad of universal Hall-type SOTs active in FM$|$NM systems. 
Therefore, AHT is an integral part of the general description of coupled spin transport and magnetization dynamics in magnetic heterostructures. 

\bigskip
\begin{large}
\noindent \textbf{\myfont Methods}
\end{large}

\noindent \textbf{\myfont Sample description.} The multilayer films were deposited by dc magnetron sputtering on Al$_2$O$_3$(0001) substrates in 2\,mTorr of Ar process gas. 
Highly resistive, amorphous Ta seed layer was used to reduce the multilayer roughness while the highly conductive Au spacer in the $\mathrm{Ta}(3\,\mathrm{nm})|\mathrm{Au}(3\,\mathrm{nm})| \NiXFe (5\,\mathrm{nm})|\mathrm{AlO_{y}}(2\,\mathrm{nm})$ multilayer was employed for enhancement of the microwave drive field applied to the $\NiFe{x}{100-x}$ magnetization in ST-FMR measurements. 
The Au spacer was removed in the $\mathrm{Ta}(5\,\mathrm{nm})| \NiFe{90}{10} (5\,\mathrm{nm})|\mathrm{AlO_{y}}(2\,\mathrm{nm})$ multilayer used for fabrication of the AHT STNO. 
Naturally oxidized AlO$_\mathrm{y}$ layer was used as a cap to prevent oxidation of $\NiXFe$. 
The multilayers were patterned into 56\,nm wide, 30\,$\mu$m long nanowires by means of electron-beam lithography using DisChem H-SiQ negative resist and Ar ion mill etching. 
The electric leads to the nanowire were patterned via electron-beam lithography using a methyl methacrylate$|$poly(methyl methacrylate) positive resist bilayer followed by deposition of $\mathrm{Ta}(5\,\mathrm{nm})| \allowbreak \mathrm{Au}(40\,\mathrm{nm})| \allowbreak \mathrm{Ta}(15\,\mathrm{nm})$ and liftoff. 
The spacing between the leads defined an active region $140-264$\,nm long in the series of $\mathrm{Ta}(3\,\mathrm{nm})| \allowbreak \mathrm{Au}(3\,\mathrm{nm})| \allowbreak \NiFe{x}{100-x} (5\,\mathrm{nm})| \allowbreak \mathrm{AlO_{y}}(2\,\mathrm{nm})$ nanowire samples. 
The active region of the $\mathrm{Ta}(5\,\mathrm{nm})| \allowbreak \NiFe{90}{10} (5\,\mathrm{nm})| \allowbreak \mathrm{AlO_{y}}(2\,\mathrm{nm})$ STNO nanowire device was was $297$\,nm long.

\noindent \textbf{\myfont Spin torque ferromagnetic resonance.} 
We employ field-modulated spin torque ferromagnetic resonance (ST-FMR) to measure the linewidths of spin wave modes as a function of $\Idc$ in the nanowire devices \cite{goncalvesSpinTorque2013}. 
All ST-FMR measurements reported in this article are made in the linear regime of magnetization dynamics. 
Application of a microwave current to the sample excites forced magnetization oscillations at the drive frequency by a combination of SOT and Oersted field from the microwave current in NM layers adjacent to the FM. 
Owing to AMR, the magnetization oscillations produce resistance oscillations at the frequency of the drive, which mix with the applied microwave current and generate the rectified voltage V$_\mathrm{mix}$. 
Resonances in V$_\mathrm{mix}$ are observed at the microwave drive frequency and applied magnetic field values corresponding to spin wave eigenmodes. 
To enhance ST-FMR signal-to-noise ratio, we employ phase-sensitive detection. 
Here, the applied magnetic field is modulated and the ac voltage $\tilde{V}_\mathrm{mix}$ is measured using a lock-in amplifier \cite{goncalvesSpinTorque2013}. 
The measured signal is proportional to magnetic field derivative of the rectified voltage $\tilde{V}_\mathrm{mix}(H)\approx\mathrm{d}V_\mathrm{mix}(H)/\mathrm{d}H$. 
The linewidth $\Delta H$ of the spin wave eigenmodes is extracted via fitting the resonances in $\tilde{V}_\mathrm{mix}(H)$ to a sum of magnetic field derivatives of the symmetric and anti-symmetric Lorentzian functions \cite{goncalvesSpinTorque2013}.
Here the linewidth $\Delta H$ is defined as half-width-at-half-maximum of the Lorentzian function.

\noindent \textbf{\myfont Microwave emission.} Microwave power spectral density generated by the AHT STNO is measured using a standard circuit based on a microwave spectrum analyser and a low-noise direct current source. 
A direct current $\Idc$ is applied to the device through the low-frequency port of a bias tee. 
The microwave signal generated by the nanowire is taken from the high-frequency port of the bias tee, amplified by a low-noise 33\,dB microwave amplifier, and recorded by the microwave spectrum analyzer. 
These measurements are made with the sample submerged in a liquid nitrogen bath $(T=77\, \mathrm{K})$. 
The values of the power spectral density reported here are those delivered to a 50$\,\Omega$ load with the frequency-dependent circuit amplification calibrated out. 

\bigskip
\begin{large}
\noindent \textbf{\myfont Data availability}
\end{large}

\noindent All data generated or analysed during this study are included in this published article and are available from the corresponding author on reasonable request.

\bibliography{refer}

\bigskip
\begin{large}
\noindent \textbf{\myfont Acknowledgements}
\end{large}

\noindent This work was supported by the National Science Foundation (ECCS-2213690).
Materials development for this project was supported by the National Science Foundation (DMREF-2324203). 
Device development and nanofabrication for this project was supported by the National Science Foundation Materials Research Science and Engineering Center program through the UC Irvine Center for Complex and Active Materials (DMR-2011967).  
The authors acknowledge the use of facilities and instrumentation at the UC Irvine Materials Research Institute (IMRI), which is supported in part by the National Science Foundation through the UC Irvine Materials Research Science and Engineering Center (DMR-2011967). 
The use of facilities and instrumentation at the Integrated Nanosystems Research Facility (INRF) in the Samueli School of Engineering at the University of California Irvine is also acknowledged.

\bigskip
\begin{large}
\noindent \textbf{\myfont Author contributions}
\end{large}

\noindent 
E.A.M. and I.N.K. designed the study. E.A.M. formulated the experimental approach. X.P. and E.A.M. deposited the multilayer samples, performed device and film level experiments, developed the nanofabrication process, and made the nano-devices. All authors wrote the manuscript and discussed the results.

\bigskip
\begin{large}
\noindent \textbf{\myfont Competing interests}
\end{large}

\noindent The authors declare no competing interests.

\end{document}


\title{\myfont Supplementary Material for: Self-generated spin-orbit torque driven by anomalous Hall current}

\title{\myfont Self-generated spin-orbit torque driven by anomalous Hall current}

\author{\myfont Eric Arturo Montoya$^{1,2,\dagger,*}$}
\author{\myfont Xinyao Pei$^{1\dagger}$}
\author{\myfont Ilya N. Krivorotov$^{1,}$}

\email{Corresponding Authors. Emails: eric.montoya@utah.edu, ilya.krivorotov@uci.edu\\$\dagger$ These authors contributed equally.}

\affiliation{\myfont $^1$ Department of Physics and Astronomy, University of California, Irvine, California 92697, USA}
\affiliation{\myfont $^2$ Department of Physics and Astronomy, University of Utah, Salt Lake City, Utah 84112, USA}

\date{\today}

\keywords{}
\maketitle

\section{Multilayer film deposition}\label{sec:films}

Multilayer films were deposited by dc magnetron sputtering on Al$_2$O$_3 (0001)$ substrates at room temperature in $2$ mTorr Ar. 
The base pressure of the system was below $3.0 \times 10^{-8} \, \mathrm{Torr}$. 
The multilayer films used in this article are composed of ferromagnetic (FM) and non-magnetic metal (NM) layers and have the structure Al$_2$O$_3$($0001$)$||$Ta($3$)$|$Au($3$)$|$Ni$_x$Fe$_{100-x}$($5$)$|$ \allowbreak AlO$_\mathrm{y}$($2$). 
Here, the numbers in parenthesis are layer thickness in nm.  
We selected a series of 5\,nm thick $\NiXFe$ alloy films as the FM layer, which allows us to tune the strength of the anomalous Hall effect (AHE) with the Ni concentration $x$ \cite{mcguireAnisotropicMagnetoresistance1975, shiEffectBand2016}. 
The NM$_1(=\mathrm{Ta}|\mathrm{Au})$ underlayer acts as an efficient spin sink while the NM$_2(=\mathrm{AlO_{y}})$ capping layer acts as a poor spin sink, thus breaking inversion symmetry. 
Additionally, the $3$ nm Ta layer is used to promote the growth of a smooth multilayer  \cite{montoyaSpinTransport2016}.  
As the NM layers remains unaltered in a series of samples with varying $x$, we will refer to the multilayer by its FM alloy composition for brevity.

The $\NiXFe$ films are grown by co-sputtering from Ni and Fe targets. 
We vary the Ni and Fe growth powers to vary alloy concentration $\NiXFe$ from $x = 65$ to $ x = 90$ and keep the FM film thickness at $\dFM = 5 \, \mathrm{nm}$. 
The growth rates are tabulated in Table {\ref{table1}}. 

\begin{table}[!hb]
\centering

\begin{tabular}{ c c  c c c }
\hline
    $x$  & Deposition rate for Ni (nm/s)  & Deposition rate for Fe (nm/s) \\
    \hline
     $65$   & $0.0108$ & $0.0058$\\
    \hline
    $70$ & $0.0117$ & $0.005$\\
    \hline
     $75$   & $0.0125$ & $0.0042$ \\
    \hline   
    $80$ & $0.0133$ & $0.0033$ \\
    \hline
    $85$  & $0.0142$ & $0.0025$ \\
    \hline
    $90$  & $0.015$ & $0.0017$ \\[1 ex]
     \hline
    
 \end{tabular}
 \caption{Deposition rates of Ni and Fe used for $\NiXFe$ film\ growth. 
 Here, $x$ is the Ni composition in Al$_2$O$_3$($0001$)$||$Ta($3$)$|$Au($3$)$|$Ni$_x$Fe$_{100-x}$($5$)$|$AlO$_\mathrm{y}$($2$). 
 The numbers in parenthesis are layer thickness in nm.}
  \label{table1}
  \end{table}

\section{Broadband ferromagnetic resonance}\label{FMR}

Film-level  ferromagnetic resonance (FMR) measurements were performed at discrete frequencies in a field swept, field modulated configuration using a broadband microwave generator, coplanar waveguide, planar-doped detector diode, and lock-in amplifier, as detailed in Ref. \cite{montoyaBroadbandFerromagnetic2014}. 
The FMR magnetization dynamics are governed by the Landau-Lifshitz-Gilbert (LLG) equation:
\begin{equation}\label{eq:LLG}
 \frac{ \partial \mathbf{M} }{ \partial t } = -\gamma \left[ \mathbf{M} \times \mathbf{H}_{\mathrm{eff}}  \right]  +  \alpha \left[ \mathbf{M} \times \frac{ \partial \uvec{m} }{ \partial t }  \right],
\end{equation}	
where $ \mathbf{M} $ is the instantaneous magnetization vector with magnitude  $ M_{\rm{s}} $, $ \uvec{m} $ is the unit vector parallel to $ \mathbf{M} $, $\mathbf{H}_{\rm{eff}}$ is the sum of internal and external magnetic fields, $\gamma = g \mu_{\mathrm{B}}/\hbar $ is the absolute value of the gyromagnetic ratio, $g$ is the Land\'{e} g-factor, $\mu_{\mathrm{B}}$ is the Bohr magneton, $\hbar$ is the reduced Planck constant, and $\alpha$ is the dimensionless Gilbert damping parameter.

\begin{figure*}[!h]
\center
 \includegraphics[width= \linewidth]{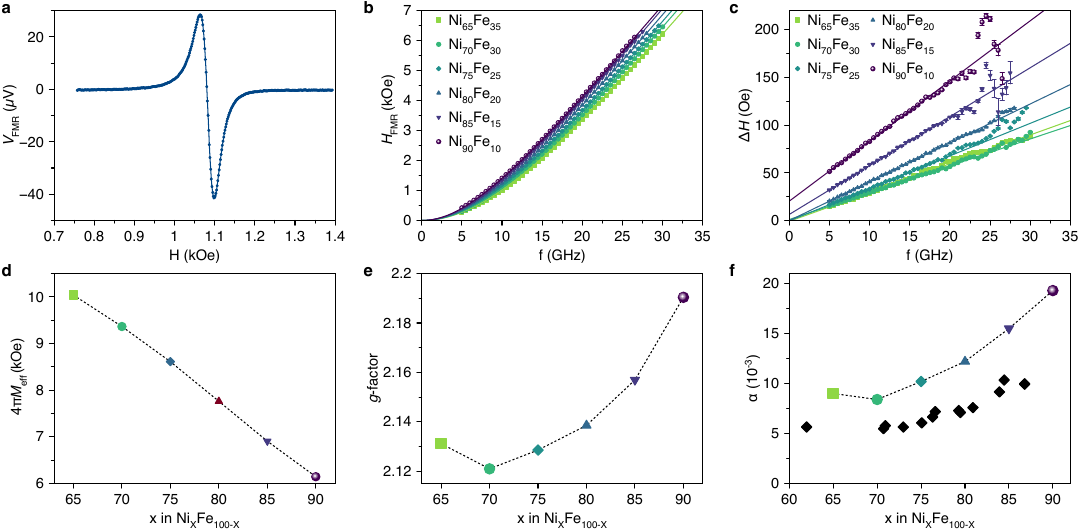}%
 \caption{\textbf{Broadband ferromagnetic resonance measurements performed on the film level.} 
 (\textbf{a}) Example of a measured FMR trace. 
 (\textbf{b}) Resonance field $\Hfmr$ as a function of frequency $f$ measured in the in-plane magnetic field configuration. 
 (\textbf{c}) FMR linewidth $\DHfmr$ as a function of frequency. 
 (\textbf{d}) Effective saturation induction $4\pi \Meff$ as a function of $x$ in $\NiXFe$. 
 (\textbf{e}) Land\'{e} g-factor  as a function of $x$ in $\NiXFe$. 
 (\textbf{f}) Gilbert damping parameter $\alpha$ as a function of $x$ in $\NiXFe$. The black diamonds indicate $\alpha$ measured in 50 nm thick $\NiXFe$ films in Ref. \cite{boninDependenceMagnetization2005}.
\label{fig:FMR_results}}%
 \end{figure*}

\Cref{fig:FMR_results}(a) shows an example FMR spectrum recorded for the $\NiFe{70}{30}$ sample (Al$_2$O$_3$ \allowbreak ($0001$)$ \allowbreak ||$Ta($3$)$|$Au($3$)$|$Ni$_{70}$Fe$_{30}$($5$)$|$AlO$_\mathrm{y}$($2$)) at $f=10 \, \mathrm{GHz}$. 
The measured FMR spectra are described by an admixture of the $\chi^{'}$ and $\chi^{''}$ components of the  complex transverse magnetic susceptibility, $ \chi = \chi^{'} + i \chi^{''} $. 
The FMR data are fit as described by Ref.~\cite{montoyaBroadbandFerromagnetic2014}. 
The $\NiXFe$ films are polycrystalline and display negligible in-plane anisotropy; therefore, the in-plane ferromagnetic resonance condition is well described by:
\begin{equation}\label{eq:Resonance}
\left( \frac{\omega}{\gamma} \right)^2 = \left( H_{\mathrm{FMR}} \right)\left( H_{\mathrm{FMR}} + 4 \pi M_{\mathrm{eff}} \right),
\end{equation}	
where $\omega =2 \pi f $ is the microwave angular frequency, $H_{\mathrm{FMR}}$ is the resonance field, and $4 \pi M_{\mathrm{eff}}$ is the effective saturation induction \cite{montoyaChapterMagnetization2015}. 
\Cref{fig:FMR_results}(b) shows the dependence of $H_{\mathrm{FMR}}$ on $f$. 
The data are fit using \cref{eq:Resonance} to extract  $4 \pi M_{\mathrm{eff}}$ and $g$; the solid lines in \cref{fig:FMR_results}(b) are the resulting fits. 
The resulting $4 \pi M_{\mathrm{eff}}$ and $g$-factors as a function of Ni concentration $x$ are plotted in \cref{fig:FMR_results}(d) and (e), respectively. 
\Cref{fig:FMR_results}(d) shows that the $4 \pi \Meff$ monotonically decreases with increasing Ni concentration in a linear fashion. 
\Cref{fig:FMR_results}(e) shows that the $g$-factor has a non-monotonic dependence on Ni concentration and has a minimum for $\NiFe{70}{30}$.

\begin{table} 
\centering

\begin{tabular}{ c c  c c c }
\hline

    Sample & $4\pi \Meff$ (Oe)  & $ g$ & $\alpha$ ($10^{-3}$)  & $\Delta H(0)$ (Oe) \\
    \hline
    Ta($3$)$|$Au($3$)$|$Ni$_{65}$Fe$_{35}$($5$)$|$AlO$_{\mathrm{y}}$($2$) & $10040$  & $2.13$ & $9.0$ & $-0.87$\\
    \hline
    Ta($3$)$|$Au($3$)$|$Ni$_{70}$Fe$_{30}$($5$)$|$AlO$_{\mathrm{y}}$($2$) & $9365$  & $2.12$ & $8.4$ & $1.00$\\
    \hline
    Ta($3$)$|$Au($3$)$|$Ni$_{75}$Fe$_{25}$($5$)$|$AlO$_{\mathrm{y}}$($2$) & $8611$  & $2.13$ & $10.2$ & $-0.42$ \\
    \hline   
    Ta($3$)$|$Au($3$)$|$Ni$_{80}$Fe$_{20}$($5$)$|$AlO$_{\mathrm{y}}$($2$) & $7762$  & $2.14$ & $12.2$ & $-0.13$ \\
    \hline
    Ta($3$)$|$Au($3$)$|$Ni$_{85}$Fe$_{15}$($5$)$|$AlO$_{\mathrm{y}}$($2$) & $6905$  & $2.16$ & $15.5$ & $6.35$ \\
    \hline
    Ta($3$)$|$Au($3$)$|$Ni$_{90}$Fe$_{10}$($5$)$|$AlO$_{\mathrm{y}}$($2$) & $6144$  & $2.19$ & $19.3$ & $20.4$ \\
    \hline
    Ta($5$)$|$Ni$_{90}$Fe$_{10}$($5$)$|$AlO$_{\mathrm{y}}$($2$) & $7308$  & $2.18$ & $14.9$ & $3.35$\\
    \hline

    \label{T2}
 \end{tabular}

 \caption{Summary of magnetic properties of multilayer films determined by broadband ferromagnetic resonance. The number in parenthesis is the layer thickness in nm. All measurements are performed at room temperature.}
  \label{table2}
  \end{table}

The measured FMR linewidth defined as half-width of the resonance curve is well described by Gilbert-like damping,
\begin{equation}\label{eq:GilbertDamping}
\Delta H(f)= \alpha \frac{2 \pi f}{\gamma} + \Delta H(0),
\end{equation}	
where $ \Delta H(0) $ is the zero-frequency line broadening due to long range magnetic inhomogeneity \cite{heinrichUltrathinMetallic1993}. 
\Cref{fig:FMR_results}(c) shows $\Delta H$ as a function of $f$ for all films. 
The data are fit using \cref{eq:GilbertDamping} to extract $\alpha$ and $\Delta H(0)$; the solid lines  are the resulting fits \cite{paikarayLargeSpin2022}. \Cref{fig:FMR_results}(f) shows the dependence of $\alpha$ on Ni concentration $x$ is non-monotonic and has a minimum for $\NiFe{70}{30}$. All magnetic parameters extracted from FMR additionally are tabulated in Table {\ref{table2}}. 

\Cref{fig:FMR_results}(f) also shows the Gilbert damping $\alpha$ in Ta(5)$|\NiXFe(50)|$Ta(5) multilayers as measured by  \citet{boninDependenceMagnetization2005}. 
Because of the fairly large (50\,nm) thickness of the FM layer and the $1/\dFM$ dependence of the damping enhancement due to spin pumping \cite{urbanGilbertDamping2001}, the spin pumping contribution to $\alpha$ in these films is negligibly small. 
The enhanced damping in our samples with Ta$|$Au and Ta NM layers compared to the $\NiXFe$(50) films in Ref.\,\cite{boninDependenceMagnetization2005} indicate that Ta$|$Au and Ta act as good spin sinks for $\NiXFe$. 
This is in agreement with previous studies demonstrating that the 3\,nm Ta layer is a very efficient spin sink for NiFe samples \cite{montoyaSpinTransport2016} and that, following the addition of the Au insertion layer, the composite Ta$|$Au layer remains a good spin sink \cite{safranskiSpinOrbit2019}. 
The AlO$_\mathrm{y}$ capping layer has been shown to act as a poor spin sink in our previous studies \cite{safranskiSpinOrbit2019}, as expected for a nonmagnetic insulator. 
Therefore, the multilayers studied here exhibit spin sinking asymmetry at the opposite FM film surfaces and thus can support an uncompensated spin accumulation at the  AlO$_\mathrm{y}$ interface as necessary to realize AHT \cite{ochoaSelfinducedSpinorbit2021a}.  

\section{Sheet Resistance}\label{Rs}

\begin{figure*}[!htb]
\center
 \includegraphics[scale=0.8]{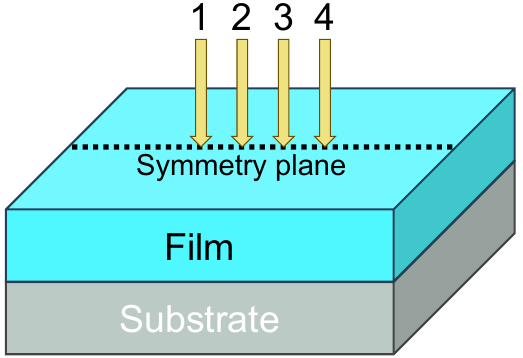}%
 \caption{\textbf{Schematic of four-point probe measurement.} 
\label{fig:4_point_probe}}%
 \end{figure*}

Sheet resistance $R_{\textrm{S}}$ is measured using a co-linear four-point probe method \cite{miccoli100thAnniversary2015}. 
Four identical probes of nominally equal spacing are placed along the symmetry plane of films and numbered in order 1-4, as shown in \cref{fig:4_point_probe}. 
The direct current $\Idc$ between points 1 and 2 is labeled as I$_{12}$ and the voltage measured at point 1 and point 3 is labeled as V$_{13}$. 

\begin{table*}[!htb]
\centering

\begin{tabular}{ c c }
\hline

    Sample & $R_{\textrm{S}}$ ($\Omega$) \\
    \hline
     Ta($3$) & $883$\\
    \hline
    Ta($3$)/Au($3$) & $49.0$\\
    \hline
    Ta($3$)$|$Au($3$)$|$Ni$_{65}$Fe$_{35}$($5$)$|$AlO$_{\mathrm{y}}$($2$) & $33.1$\\
    \hline   
    Ta($3$)$|$Au($3$)$|$Ni$_{70}$Fe$_{30}$($5$)$|$AlO$_{\mathrm{y}}$($2$) & $32.8$\\
    \hline
    Ta($3$)$|$Au($3$)$|$Ni$_{75}$Fe$_{25}$($5$)$|$AlO$_{\mathrm{y}}$($2$)  & $30.9$\\
    \hline
    Ta($3$)$|$Au($3$)$|$Ni$_{80}$Fe$_{20}$($5$)$|$AlO$_{\mathrm{y}}$($2$)  & $30.4$\\
    \hline
    Ta($3$)$|$Au($3$)$|$Ni$_{85}$Fe$_{15}$($5$)$|$AlO$_{\mathrm{y}}$($2$)  & $28.9$\\
    \hline
    Ta($3$)$|$Au($3$)$|$Ni$_{90}$Fe$_{10}$($5$)$|$AlO$_{\mathrm{y}}$($2$)  & $28.0$\\
    \hline 
    Ta($5$) & $450$\\
    \hline
    Ta($5$)$|$Ni$_{90}$Fe$_{10}$($5$)$|$AlO$_{\mathrm{y}}$($2$)  & $57.6$\\
    \hline    
 \end{tabular}
 \caption{ \textbf{Sheet resistance determined by four-point probe measurements.} Numbers in parenthesis are layer thicknesses in nm. All measurements are performed at room temperature.}
 \label{tableRs}
  \end{table*}

The resistances $R_A=\frac{V_{23}}{I_{14}}$,
$R_{B}=\frac{V_{24}}{I_{13}}$, and  $R_{C}=\frac{V_{43}}{I_{12}}$ are measured using a four terminal source-meter and R$_A$, R$_B$ and R$_C$ satisfy \cref{s1}) and (\cref{s2} \cite{miccoli100thAnniversary2015}:
\begin{equation}
\label{s1}
e^{-\frac{2\pi R_{A}}{R_{\textrm{S}}}}+e^{-\frac{2\pi R_{C}}{R_{\textrm{S}}}}=1
\end{equation}
\begin{equation}
\label{s2}
e^{-\frac{2\pi R_{A}}{R_{\textrm{S}}}}+e^{-\frac{2\pi R_{B}}{R_{\textrm{S}}}}=1.
\end{equation}
\noindent The symmetric measurement for the $R_{\textrm{S}}$ is averaged and the voltage leads are switched to correct for small deviations in probe spacing \cite{miccoli100thAnniversary2015}. Here, we have used:
\begin{equation}
R_{A}=(\frac{V_{23}}{I_{14}}+\frac{\vert V_{32}\vert}{I_{14}})/2 
\end{equation}
\begin{equation}
R_{B}=(\frac{V_{24}}{I_{13}}+\frac{V_{13}}{I_{24}}+\frac{\vert V_{42}\vert}{I_{13}}+\frac{\vert V_{31}\vert}{I_{24}})/4
\end{equation}
\begin{equation}
R_{C}=(\frac{V_{43}}{I_{12}}+\frac{V_{12}}{I_{43}}+\frac{\vert V_{34}\vert}{I_{12}}+\frac{\vert V_{21}\vert}{I_{43}})/4\\ 
\end{equation}

The result of the $R_{\textrm{S}}$ measurements is summarized in Table \ref{tableRs}. The data in Table \ref{tableRs} shows that as the alloy concentration $x$ of Ni increases, $R_{\textrm{S}}$ decreases.

\section{Frequency dependence of anomalous Hall torque}\label{sec:freqAHT}

\begin{figure}[htpb]
\center
 \includegraphics{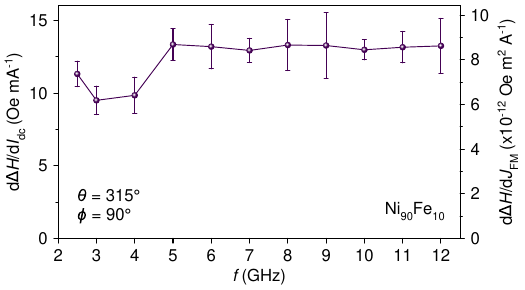}%
 \caption{\textbf{Frequency dependence of anomalous Hall torque for the Ta$|$Au$|$Ni$_{90}$Fe$_{10}|$AlO$_{\mathrm{y}}$ nanowire device.}  
\label{fig:frequency_dependence_AHT}}%
 \end{figure}

According to the theory of \citet{ochoaSelfinducedSpinorbit2021a}, the current-induced change in the resonance linewidth, proportional to the magnitude of the self-generated anomalous Hall torque, is independent of drive frequency and applied magnetic field, see equation (14) in \cite{ochoaSelfinducedSpinorbit2021a}. 
\Cref{fig:frequency_dependence_AHT} shows $d\Delta H / d \Idc$ as a function of frequency for the Ta$|$Au$|$Ni$_{90}$Fe$_{10}|$AlO$_{\mathrm{y}}$ nanowire with magnetic field applied at  $\theta=315^\circ$ and $\phi = 90^\circ$. We find that the AHT is independent of frequency and applied magnetic field strength for frequencies above 5\,GHz for this sample. 
We attribute the deviation below 5\,GHz to the applied field not being strong enough to saturate the magnetization vector parallel to magnetic field vector. In such a case, the magnetization vector is actually closer to being in the plane of the sample with a larger component along the easy $x$-axis and thus we observe a reduction in the measured AHT strength.

\section{Magnetotransport Measurements}\label{AHE}

\begin{figure}[!htb]
\center
 \includegraphics[width= 0.95\linewidth]{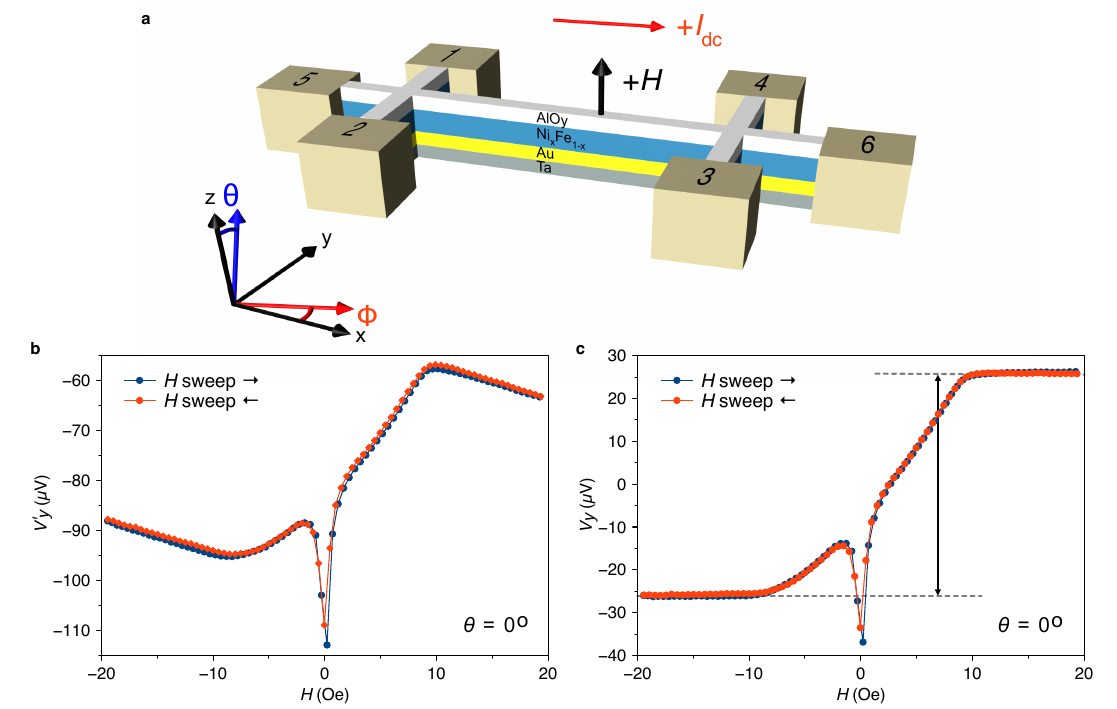}%
 \caption{\textbf{Anomalous Hall effect measurements.} 
 \textbf{(a)} Schematic of the Hall bar with coordinate system and contact labels. 
 \textbf{(b)} Total Hall voltage $V_y'$ as function of field for Ta$|$Au$|$Ni$_{70}$Fe$_{30}|$AlO$_{\mathrm{y}}$. 
 \textbf{(c)} Anomalous Hall voltage $V_y$ as function of field for Ta$|$Au$|$Ni$_{70}$Fe$_{30}|$AlO$_{\mathrm{y}}$ after subtracting out ordinary Hall effect.
\label{fig:Hallbar}}%
 \end{figure}

We measure anomalous Hall charge resistivity $\rhoAHE$ using 6-contact Hall bars deposited under the same conditions as the films used to fabricate the nanowires. 
The Hall bars are fabricated using photolithography and lift-off techniques.  \Cref{fig:Hallbar}(a) shows a schematic of a Hall bar. 
The Hall bar is placed in a magnetic field oriented along the surface normal ($z$-axis) with positive field corresponding to the  $+z$-direction. 
A low-noise current source supplies a bias current $\Idc = 0.1 \, \mathrm{mA} $  between contacts 5 and 6 such that positive conventional current flows along the $x$-axis by entering contact 5 ($I^+$ terminal) and exiting contact 6 ($I^-$ terminal). 
A nanovoltmeter is connected to contacts 3 and 4 with positive terminal connected to 3 ($V^{+}$) and negative terminal connected to 4 ($V^{-}$). 
The Hall voltage $V_y^{'} = V_3 -V_4$ is measured as a function of applied magnetic field $H$. \Cref{fig:Hallbar}(b) shows $V_y^{'}$ for the Ta$|$Au$|$\NiFe{70}{30}$|$AlO$_\mathrm{y}$ Hall bar displaying both anomalous Hall effect and ordinary Hall effect. 

\Cref{fig:Hallbar} (c) shows the anomalous Hall charge voltage $V_y$ obtained after removal of the ordinary Hall contribution. 
At zero field  the magnetization lies primarily along the easy $-x$-direction and AHE does not contribute to the transverse voltage $V_{y}$. 
Here there is a background contribution due to planar Hall effect (PHE). 
As the perpendicular-to-plane field is increased, the magnetization is pulled toward the $z$-axis, increasing the magnitude of $V_{y}$ until the magnetization is saturated (near 10 kOe here). 
For positive field in $z$-direction, the saturation voltage is $V_{y}^{+}$, while for negative field it is $V_{y}^{-}$. 
The anomalous Hall charge voltage is defined as $ \VAHE = \left( V_{y}^{+} - V_{y}^{-} \right)/2$. 

We note that sign conventions pertaining to Hall measurements and AHE vary in literature. 
We use the convention that Hall charge effects at saturation in an isolated FM layer can be expressed as \cite{smitSpontaneousHall1953}:
\begin{equation}\label{eq:HV}
    \frac{E_y^{'}}{J_{x}} = A_{\mathrm{H}}H_{z} + \rhoAHE,
\end{equation}
where $J_x$ is the charge current density, $A_\mathrm{H}$ is the ordinary Hall coefficient, and $H_z$ is the applied saturating field. 

\begin{figure}
\includegraphics[width= \linewidth]{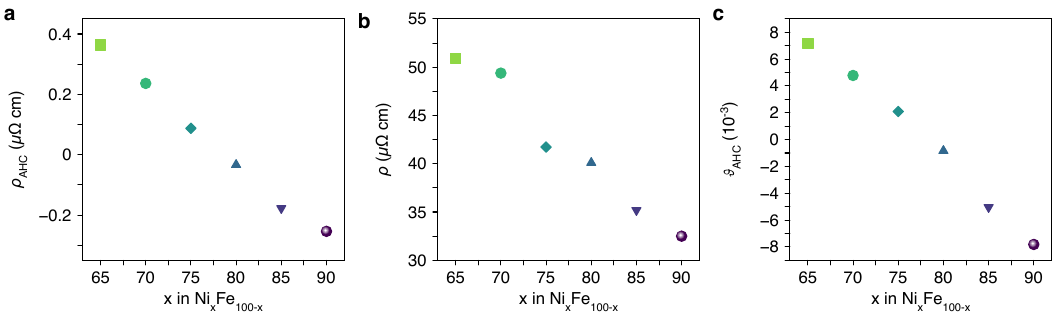}%
 \caption{\textbf{Transport measurements.} 
 \textbf{(a)} Anomalous Hall charge resistivity as function of $x$ in Ta$|$Au$|$Ni$_{x}$Fe$_{100-x}|$AlO$_{\mathrm{y}}$. 
 \textbf{(b)} Resistivity of Ni$_{x}$Fe$_{100-x}$ layer as determined from $4-$point-probe measurements. 
 \textbf{(c)} Anomalous Hall angle in Ni$_{x}$Fe$_{100-x}$. 
\label{fig:Res}}%
 \end{figure}

However, our samples are bilayers consisting of FM layer and NM layers. 
To convert the measured $ \VAHE$ to the anomalous Hall charge resistivity $ \rhoAHE = \rhoyxFM = -\rhoxyFM $, we must consider the effect of counter flow charge current in the adjacent NM layers. 
In the geometry of \cref{fig:Hallbar}(a), the transverse charge current density induced by AHE in the FM layer is:
\begin{equation}
    \jyFM = \sigmayxFM \Ex = - \sigmaxyFM \Ex.
\end{equation}
Since the net transverse charge current in the bilayer must be zero, there must be a transverse counter-flow current $\Iycf$ so that
\begin{equation}
    \Iynet = \IyFM + \Iycf = 0.
\end{equation}
Here the currents are transverse currents per unit length of the wire. For example:
\begin{equation}
    \IyFM = \jyFM \dFM,
\end{equation}
where $\dFM$ is the FM layer thickness.

Since the counter-flow current flows through both FM and NM layers, the condition of zero net transverse current becomes:
\begin{equation}
    \Iynet = \jyFM \dFM + \jyFMcf \dFM + \jyNMcf \dNM = 0,
\end{equation}
where $\dNM$ is the NM layer thickness while $\jyFMcf$ and $\jyNMcf$  are counter-flow charge current densities in the FM and NM layers, respectively. 
Expressing current densities in terms of conductivities:
\begin{equation}
    \sigmayxFM \Ex \dFM + \sigmayyFM \Ey \dFM + \sigmayyNM \Ey \dNM = 0.
\end{equation}
Here $\sigmayyFM$ and $\sigmayyNM$ are longitudinal conductivities of the FM and NM layers, respectively, while $\Ey$ is the transverse electric field from the charge accumulation at the edges of the wire. Solving this equation for $\Ey$, we obtain: 
\begin{equation}
    \Ey = \frac{\sigmaxyFM \dFM}{\sigmayyFM \dFM + \sigmayyNM \dNM} \Ex.
\end{equation}
If the net charge current flowing along the wire is $\Idc$, then $\Ex$ can be expressed in terms of this current:
\begin{equation}
    \Ex = \frac{1}{w} \frac{\Idc}{\sigmaxxFM \dFM + \sigmaxxNM \dNM},
\end{equation}
where $w$ is the width of the nanowire in the $y$-axis. Substitution in the expression for $\Ey$ and assuming longitudinal conductivities along $x$- and $y$-axes are identical for both FM and NM  ($\sigmaxxFM = \sigmayyFM$ and $\sigmaxxNM = \sigmayyNM$):
\begin{equation}
    \Ey = \frac{1}{w} \frac{\sigmaxyFM \dFM}{\left( \sigmaxxFM \dFM + \sigmaxxNM \dNM \right)^2} \Idc.
\end{equation}

\begin{figure}[!htb]
\center
 \includegraphics{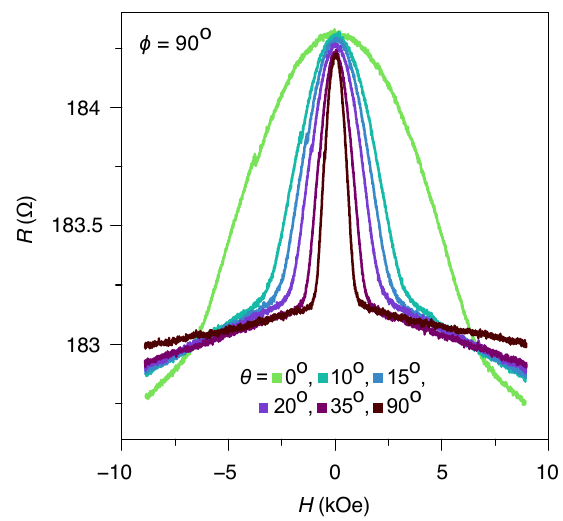}%
 \caption{\textbf{Magnetoresistance measurement.} Resistance versus magnetic field measured for the Ta$|$Au$|$Ni$_{70}$Fe$_{30}|$AlO$_{\mathrm{y}}$ nanowire at various magnetic field directions.
\label{fig:RvsH}}%
 \end{figure}

Therefore, measured anomalous Hall voltage in the bilayer $\Vy$ is:
\begin{equation}
    \Vy = \Ey w = \frac{\sigmaxyFM \dFM}{\left( \sigmaxxFM \dFM + \sigmaxxNM \dNM \right)^2} \Idc.
\end{equation}
Taking,
\begin{equation}
    \rhoyxFM = \frac{\sigmaxyFM}{\left( \sigmaxxFM \right)^2},
\end{equation}
we derive:
\begin{equation}
    \rhoyxFM = \frac{\left( \dFM + \dNM \left( \sigmaxxNM / \sigmaxxFM \right) \right)^2}{\dFM } \frac{\Vy}{\Idc}.
\end{equation}
The solution can be extended to two NM layers (NM$_1$ and NM$_2$) as:
\begin{equation}
    \rhoyxFM = \frac{\left( \dFM + d_\mathrm{NM1} \left( \sigma_\mathrm{xx}^{\mathrm{NM}1} / \sigmaxxFM  \right) + d_\mathrm{NM2} \left( \sigma_\mathrm{xx}^{\mathrm{NM}2} / \sigmaxxFM \right) \right)^2 }{\dFM } \frac{\Vy}{\Idc}.
\end{equation}
We then define
\begin{equation}
    \rhoAHE \equiv \rhoyxFM.
\end{equation}

\Cref{fig:Res}(a) shows $\rhoAHE$ is strongly dependent on $x$, changing sign between $x$ $=75$ and $80$. \Cref{fig:Res}(b) shows the resistivity of the FM layer determined using $4$-point-probe measurements decreases with the increasing Ni concentration. \Cref{fig:Res}(c) shows the anomalous Hall charge angle $\thetaAHE \equiv \rhoAHE / \rho$.
These results are in good agreement with prior measurements \cite{mcguireAnisotropicMagnetoresistance1975, shiEffectBand2016}.

\Cref{fig:RvsH} shows the sizable AMR for the Ta$|$Au$|$Ni$_{70}$Fe$_{30}|$AlO$_{\mathrm{y}}$ nanowire device used for ST-FMR measurements in the main text. 

\section{Anomalous Hall torque model}\label{AHT}

\textbf{Two-current model approximation.} To build an intuitive physical picture of the anomalous Hall torque origin, we first use the two-current spin–majority and spin–minority model. 
This is only an approximation because it neglects spin state mixing by spin-orbit coupling (SOC). 
This is not a bad qualitative approximation for the materials where SOC fairly small, but it should not be used for quantitative calculations. 
Within the two-channel current model, we monitor the motion of spin-majority and spin-minority carriers in the ferromagnetic (FM) conductor.

\begin{figure*}[htpb]
\center
 \includegraphics[width= 0.5\linewidth]{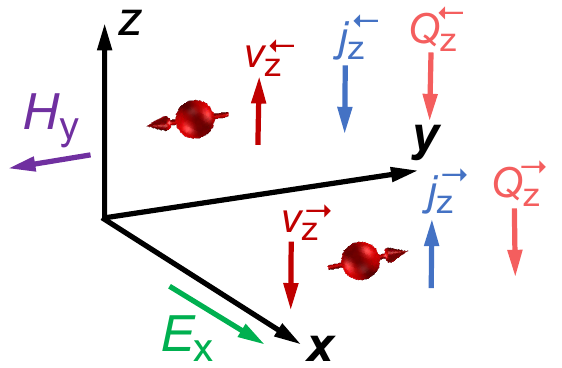}%
 \caption{\textbf{Transverse charge and spin transport in an infinite ferromagnet.} The infinite FM model shows transverse motion of electrons with directions of the electron's magnetic moment, spin current densities and conductivities. 
\label{fig:transport_infinite_FM}}%
 \end{figure*}

We first examine an infinite FM. When electric field is applied in the $+x$-direction (and the longitudinal electric current flows in the $+x$-direction) and magnetic field $H$ is applied in the $-y$-direction as shown in \cref{fig:transport_infinite_FM}, electrons acquire a velocity component transverse to the $x$-axis owing to the Hall effects.
Here $\vleft$ is transverse velocity of the majority electrons while $\vright$ is transverse velocity of the minority electrons (we use the $\vleft$ and $\vright$ instead of the customary $\vup$ and $\vdown$ notations to better visualize the in-plane spin directions in the figures of this section). 
The red arrows attached to the electrons (red spheres) show the directions of the electron’s magnetic moment (spin of these electrons points in the opposite direction to their magnetic moments).

There are transverse electric current densities ($\jzleft$ and $\jzright$) and conductivities ($\sigmazxleft$ and $\sigmazxright$) associated with this transverse motion of electrons (the electrical conductivity tensor is antisymmetric): 

\begin{equation}
    \jzleft = \sigmazxleft \Ex = -\sigmaxzleft \Ex
\end{equation}
\begin{equation}
    \jzright = \sigmazxright \Ex = -\sigmaxzright \Ex
\end{equation}

The current densities and conductivities depend on the $y$-component of magnetization. 
For simplicity of the discussion, we assume that magnetization is saturated along the $-y-$direction.  
The transverse conductivities can be either positive or negative. For the direction of motion of the electrons shown in \cref{fig:transport_infinite_FM}, $\vleft_z > 0 $ corresponds to $\sigmaxzleft> 0$  (electron charge is negative so current flows in the direction opposite to the electron flow) and $\vright_z < 0$ corresponds to $\sigmaxzright < 0$.

The net transverse charge current density is a sum of the transverse charge current densities of the majority and minority electrons: 

\begin{equation}
    \jz = \jzleft + \jzright = \left( \sigmazxleft + \sigmazxright \right)\Ex = - \left( \sigmaxzleft + \sigmaxzright \right) \Ex
\end{equation}

The definition of the anomalous Hall charge conductivity $\sigmaxzAHE$ is:
\begin{equation}
    \jz = \sigmazxAHE \Ex = - \sigmaxzAHE \Ex
\end{equation}
Therefore, in the two-channel current model:
\begin{equation}
    \sigmaxzAHE = \left( \sigmaxzleft + \sigmaxzright \right)
\end{equation}

Coupled to the transverse electron velocity in this model, there is a transverse spin current density $\Qzleft$ and $\Qzright$ (spin angular momentum current density to be more precise). 
This spin current density flows along the $z$-axis and the spins of these spin currents are polarized along the $y$-axis in our example.

\begin{equation}\label{eq:Qzleft}
    \Qzleft = \frac{\hbar}{2 |e|} \sigmazxleft \Ex = - \frac{\hbar}{2 |e|} \sigmaxzleft \Ex
\end{equation}
\begin{equation}\label{eq:Qzright}
    \Qzright = - \frac{\hbar}{2 |e|} \sigmazxright \Ex = \frac{\hbar}{2 |e|} \sigmaxzright \Ex
\end{equation}
\Cref{eq:Qzleft,eq:Qzright} reflect the fact that each electron of electric charge $e$ carries a $y$-axis spin angular momentum of $\pm \hbar/2$. Notice the sign difference between the majority and minority electron expressions. This sign difference is due to the spin angular momentum reversal between these two cases (reversal of spin leads to reversal of spin current direction).

To develop better intuition about these spin current densities:
\begin{enumerate}
    \item $\Qzleft < 0 \left( \sigmaxzleft > 0 \right)$ means that positive $\left(  +y \right)$ spin angular momentum (majority electrons) flows in the $+z$-direction $\left( \vleft>0 \right)$ 
    \item $\Qzright < 0 \left( \sigmaxzright < 0 \right)$ means that negative $\left(  -y \right)$ spin angular momentum (minority electrons) flows in the $-z$-direction $\left( \vright<0 \right)$ 
\end{enumerate}

\Cref{fig:transport_infinite_FM} illustrates the directions of spin current densities corresponding to the directions of transverse electron velocities shown.
\begin{figure*}[htpb]
\center
 \includegraphics[width= 0.65\linewidth]{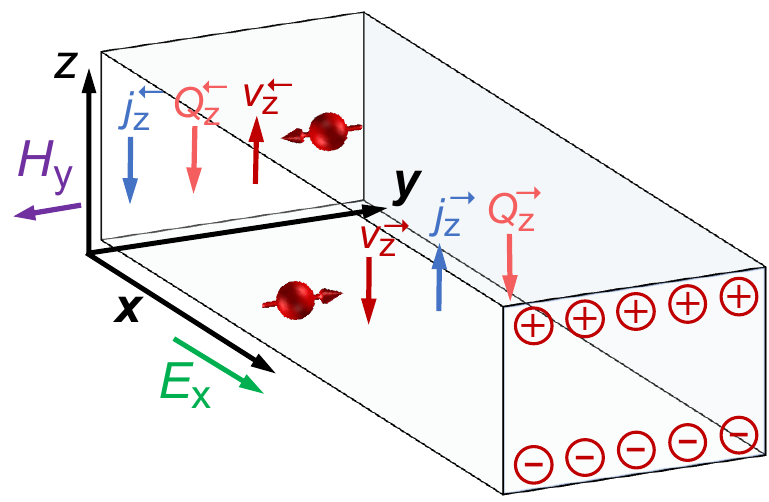}%
 \caption{\textbf{Charge accumulation in a ferromagnetic film.}  The FM film model shows transverse motion of electrons with directions of the electron's magnetic moment, spin current densities and conductivities. Charge accumulation illustrated for Ni-like system with anomalous Hall conductivity $\sigmaxzAHE<0$.
\label{fig:charge_accumulation}}%
 \end{figure*}

The net spin angular momentum current along the $z$-axis is:
\begin{equation}
    \Qz = \Qzleft + \Qzright =  - \frac{\hbar}{2 |e|} \left( \sigmaxzleft - \sigmaxzright \right) \Ex
    \label{eq:S30}
\end{equation}
$ \Qz >0 $ means that either negative $\left( -y \right)$ spin angular momentum flows in the $+z$-direction or positive $\left( +y \right)$ spin angular momentum flows in the $-z$-direction.

Now let us consider the case of FM nanowire instead of an infinite FM. If $\sigmaxzAHE$ of the material is non-zero, then charge accumulates at the wire surfaces along the $z$-axis (at the top and bottom of the FM film). 
For concreteness, let us assume that $\sigmaxzleft > 0$, $\sigmaxzright < 0 $, and $\left| \sigmaxzleft \right| < \left| \sigmaxzright\right|$ -- this is the case for Ni \cite{aminIntrinsicSpin2019} in the two-channel approximation. 
This means that $ \sigmaxzAHE < 0 $ and negative charges accumulate on the $-z$-surface of the FM wire as shown in \Cref{fig:charge_accumulation}.

Due to the finite extent of the sample along the $z$-axis, net charge current in the $z$-direction must be zero in the steady state. 
We thus must have a counter-flow charge current density in the $z$-axis $\jzcf$ that is driven by the surface charge accumulation and cancels the transverse anomalous Hall charge current:
\begin{equation}
    \jz + \jzcf = 0
\end{equation}
\begin{equation}
    \jzcf = \sigmaxzAHE \Ex
\end{equation}
This counter-flow charge current is spin-polarized. 
Since this charge current is driven by the charge accumulation rather than Hall effects, its spin polarization is the same as that of the longitudinal charge current in this FM material. 
Polarization of the longitudinal current $P$ is defined as \cite{ochoaSelfinducedSpinorbit2021a, zhu_CurrentPolarization_2010, kokadoAnisotropicMagnetoresistance2012}:
\begin{equation}\label{eq:Polarization}
    P = \frac{\sigmaxxleft - \sigmaxxright}{\sigmaxxleft + \sigmaxxright}
\end{equation}
In Ni, the longitudinal charge current is carried predominantly by the majority electrons, so $\sigmaxxup > \sigmaxxdown$ , so $P > 0$ \cite{kokadoAnisotropicMagnetoresistance2012}. Thus, the counter-flow charge current carries spin current density:
\begin{equation}\label{eq:Qzcf}
    \Qzcf = \frac{\hbar}{2 |e|} P \sigmaxzAHE \Ex
\end{equation}

Similar to $\Qz$ described by \cref{eq:S30}, $\Qzcf>0$ means that either negative ($-y$) spin angular momentum flows in the $+z$-direction or positive ($+y$) spin angular momentum flows in the $-z$-direction.

\begin{figure*}[htpb]
\center
 \includegraphics[width= 0.65\linewidth]{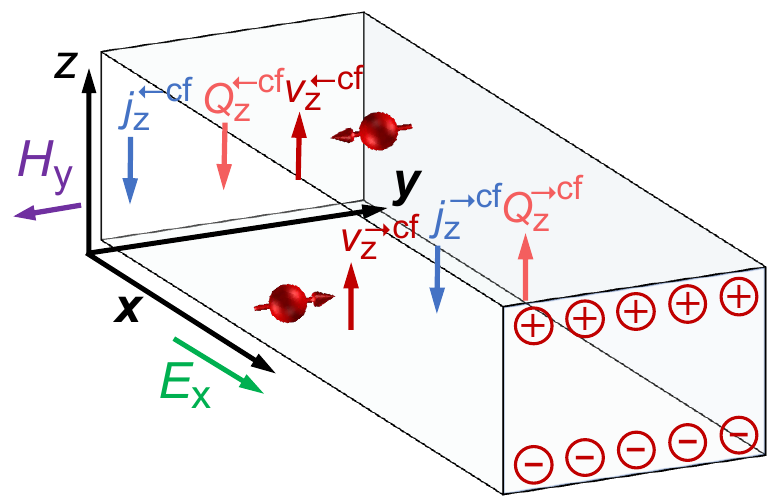}%
 \caption{\textbf{Counter flow current in  a ferromagnetic film.} The FM film model shows directions of counter-flow charge and spin current densities. Charge accumulation and backflow polarization illustrated for Ni-like system with anomalous Hall conductivity $\sigmaxzAHE<0$ and magnetic polarization $P>0$.
\label{fig:definitions_wire_cf}}%
 \end{figure*}

\Cref{fig:definitions_wire_cf} illustrates the counter-flow charge and spin current densities resolved by spin current channel.
The net spin current density is:
\begin{equation}
    \Qznet = \Qz + \Qzcf = -\frac{\hbar}{2 |e|} \left( \sigmaxzleft - \sigmaxzright \right) \Ex + \frac{\hbar}{2 |e|} P \sigmaxzAHE \Ex
\end{equation}

\bigskip

\textbf{Beyond the two-current model.} Now we can go beyond the two-current model. In a more realistic picture, the current cannot be represented by a sum currents in two independent spin channels because these channels are mixed by SOC. 
Nevertheless, there is a transverse spin current density with its spin polarization collinear with magnetization \cite{das_efficient_2018, aminIntrinsicSpin2019,miuraFirstprinciplesCalculations2021}. 
Similar to \cref{eq:S30}, it can be written as:
\begin{equation}
    \Qz =  -\frac{\hbar}{2 |e|} \sigmaxzX \Ex
\end{equation}
where $\sigmaxzX$ is transverse anomalous Hall spin conductivity (sum of spin anomalous Hall conductivity $\sigmaxySAHE$ and spin Hall conductivity $\sigmaxySHE$ as in Refs. \cite{aminIntrinsicSpin2019,miuraFirstprinciplesCalculations2021}, all expressed in the units of charge conductivity). The net transverse anomalous Hall spin  current density with its polarization collinear with magnetization in this general case is:
\begin{equation}
    \Qznet =  -\frac{\hbar}{2 |e|} \sigmaxzX \Ex + \frac{\hbar}{2 |e|} P \sigmaxzAHE \Ex
\end{equation}
\begin{equation}
    \Qznet =  -\frac{\hbar}{2 |e|} \left( \sigmaxzX  - P \sigmaxzAHE \right) \Ex
    \label{eq:S38}
\end{equation}
The right hand side of this equation can be rewritten as \cite{ochoaSelfinducedSpinorbit2021a}:
\begin{equation}
    \Qznet = -\frac{\hbar}{2 |e|} \sigmaxx \left(  \thetaX  - P \thetaAHE \right) \Ex
    \label{eq:S39}
\end{equation}
where $\sigmaxx$ is the longitudinal charge conductivity of the FM.
The damping-like AHT is proportional to $\Qznet$.

According to \citet{aminIntrinsicSpin2019}, $\thetaX >0$  and $\thetaAHE < 0 $ for Ni. 
For for $\Ex > 0$ and $ H_\mathrm{y} < 0$ as illustrated in \cref{fig:definitions_wire_cf}, this means that the transverse anomalous Hall spin current in Ni is negative. 
Therefore  spins with $+y$-direction (majority-like) accumulate on the same surface ($+z$-surface) where positive charge accumulates while spins with $-y$-direction (minority-like) accumulate on the $-z$-surface.  
To establish correspondence to our nanowire geometry, imagine that the poor spin sink is at the $+z$-surface while the good spin sink is at the $-z$-surface and $H_\mathrm{y} < 0$. 
The good spin sink prevents spin accumulation at the $-z$-surface and we are left with  spin accumulation in the $+y$-direction  (majority-like) at the $+z$-surface. 
In our Ni-rich nanowires, this corresponds to the net current-driven enhancement of magnetization in the wire (for $\Ex > 0$ and $H_\mathrm{y} < 0$) and thus increase of the effective damping. 
Under magnetization reversal ($\Ex > 0$ and $H_\mathrm{y} > 0$), the spin accumulation at the $+z$-surface is in the $y$-direction (minority-like), leading to a reduction of the magnetization in the wire and thus a decrease in the damping (antidamping). 
This is what we observe in our experiment.

\section{Calculation of the AHT-induced damping efficiency \texorpdfstring{$A_\mathrm{yz}$}{Ayz}}\label{sec:Ayz}

Here we derive an  expression for the critical current density \cite{kiselevMicrowaveOscillations2003, rippard_contact_STO_2004} of the AHT-based spin torque nano-oscillator (STNO) and efficiency of AHT-induced damping $A_\mathrm{yz}$ using the general theory of STNO \cite{slavinNonlinearAutoOscillator2009} and the AHT theory \cite{ochoaSelfinducedSpinorbit2021a}. 
The Gilbert damping torque $\mathbf{T}_{\mathrm{G}}$ is:
\begin{equation}
    \mathbf{T}_{\mathrm{G}}  
    = \gamma \alpha M_\mathrm{s} \left[ \mathbf{m} \times \left[  \mathbf{H}_\mathrm{eff} \times \mathbf{m} \right] \right].
    \label{eq1}
\end{equation}
For comparison, we first derive the expression for critical current density of a SHT-based STNO \cite{ranjbar_cofeb-basedSHO_2014, demidovSpinOrbittorque2020, smithDimensionalCrossover2020}. Spin Hall torque $\mathbf{T}_{\mathrm{SHT}}$ can be written as:
\begin{equation}
    \mathbf{T}_{\mathrm{SHT}} = \gamma\frac{\hbar}{2 |e| }\frac{\vartheta_\mathrm{SHE} J_{\mathrm{NM}}}{\dFM }\left[ \mathbf{m} \times \left[  \mathbf{m} \times \hat{\mathbf{y}} \right] \right],
    \label{eq5}
\end{equation}
where $e$ is electron charge, $g$ is Land\'{e} g-factor, $\mu_\mathrm{B}$ is the electron Bohr magneton, $\dFM$ is the FM layer thickness, $J_{\mathrm{NM}}$ is charge current density in  the NM layer of the FM$|$NM bilayer and $\vartheta_\mathrm{SHE}$ is spin Hall angle of the NM material. 
Spin polarization of the spin Hall current is along the $y$-axis and $\hat{\mathbf{y}}$ is a unit vector in the $+y$-direction. We note that \cref{eq5} is defined for bilayer where the spin Hall material is on the FM film bottom. For spin Hall material on top the FM film, the sign of \cref{eq5} must be changed.

Making a simplifying assumption of circular magnetic precession and assuming that the equilibrium direction of magnetization lies in the $yz$-plane, we can calculate the critical current density for a ST-based STNO $J_\mathrm{NM0}$ from \cref{eq1} to \cref{eq5}. 
Only the component of spin current polarization collinear with $\mathbf{H}_{\mathrm{eff}}$ contributes to the current-induced antidamping and compensates the Gilbert damping torque at the critical current density. 
This damping compensation condition gives:
\begin{equation}  \gamma\alpha M_{\mathrm{s}} H_{\mathrm{eff}} = \gamma\frac{\hbar}{2 |e| }\frac{\vartheta_\mathrm{SHE} J_{\mathrm{NM0}}}{\dFM }\sin(\theta),
\end{equation}
where $\theta$ is the polar angle between the $z$-axis and $\mathbf{H}_{\mathrm{eff}}$ in the $yz$-plane ($\phi = 90^{\circ}$). 
Solving for $ J_{\mathrm{NM0}}$:
\begin{equation}
    J_{\mathrm{NM0}}= \frac{2 |e|}{\hbar}
    \frac{\alpha}{\vartheta_\mathrm{SHE}}
    \frac{M_{\mathrm{s}} H_{\mathrm{eff}}\dFM }{\sin(\theta)}.
    \label{eq7}
\end{equation}
\Cref{eq7} can be written in several equivalent forms using the following identities: $\omega=\gamma H_{\mathrm{eff}}$, $\Delta H=\alpha H_{\mathrm{eff}}$, and $\gamma=g \mu_{\mathrm{B}}/\hbar$: 
\begin{equation}
    J_{\mathrm{NM0}}= \frac{2 |e|}{\hbar}
    \frac{\dFM M_{\mathrm{s}}\Delta H}{\vartheta_\mathrm{SHE}   \sin(\theta)}
\end{equation}
or
\begin{equation}
    J_{\mathrm{NM0}}=2 |e|\omega \frac{ M_{\mathrm{s}} \dFM}{g\mu_{\mathrm{B}}}\frac{\alpha}{\vartheta_\mathrm{SHE} \sin(\theta)}
    \label{eq9}
\end{equation}
\Cref{eq9} is particularly convenient for predicting the critical current density of spin Hall oscillators.
 
The expression for the AHT vector was derived in \cite{ochoaSelfinducedSpinorbit2021a}. 
Here we use this expression with an additional assumption that the angular momentum carried by spin current $Q^\mathrm{net}_\mathrm{z}$ is converted to AHT with the maximum efficiency allowed by symmetry of the system (this corresponds to the ratio of torque efficiency to spin relaxation rate $\eta/\Gamma_\mathrm{s}=1$ in the notations of
of Ref. \cite{ochoaSelfinducedSpinorbit2021a}):
\begin{equation}\label{eq:TorqeeFormAHT1}
\mathbf{T}_\mathrm{AHT}=  -\gamma \frac{\hbar}{2 |e|} \frac{\thetaAHT}{d_{FM}} m_z (\bf{m} \times \hat{z} \times \bf{m})(\bf{m} \cdot (\hat{z} \times \bf{\JFM})),
\end{equation}
where $\bf{\JFM}$ is charge current density in the FM. If additional angular momentum channels are present, the right hand side of \cref{eq:TorqeeFormAHT1} should be multiplied by a dimensionless parameter $<1$ quantifying the loss. We note that \cref{eq:TorqeeFormAHT1} is defined for bilayer where the good spin sink is on the FM film bottom and poor spin sink is on film top, as the case for our devices. For good spin sink on top and poor spin sink on the bottom of the FM film, the sign of \cref{eq:TorqeeFormAHT1} must be changed.

This AHT acts as antidamping torque and, at the critical current density $J_\mathrm{FM0}$, it cancels the Gilbert damping torque \cite{ochoaSelfinducedSpinorbit2021a}. This cancellation condition gives the critical current density for magnetization in the $yz$-plane ($\phi=90^\circ$):
\begin{equation}
    J_{\mathrm{FM0}}=2 |e|\omega \frac{ M_{\mathrm{s}} \dFM}{g\mu_{\mathrm{B}}}\frac{\alpha}{\vartheta_\mathrm{AHT} \sin(\theta)\cos^2(\theta)}
    \label{eq:AHT_critital_current}
\end{equation}
This expression is similar in form to that for the SHT critical current given by \cref{eq9}. The differences between \cref{eq9} and \cref{eq:AHT_critital_current} are in the angular dependence and the dimensionless parameter governing the torque magnitude ($\vartheta_\mathrm{SHT}$ versus $\vartheta_\mathrm{AHT}$).
\Cref{eq:AHT_critital_current} can be rewritten in a slightly different form by using the trigonometric identity $\sin(\theta)\cos^2(\theta)=\frac{1}{4}(\sin(\theta)+\sin(3\theta))$. 

\Cref{eq:AHT_critital_current} and Eq.\,(1) of the main text can be used to calculate the AHT-induced damping efficiency parameter $\Ayz$:
\begin{equation}\label{eq:Ayz}
    \Ayz = \frac{\hbar}{8 |e|}\frac{\thetaAHT}{M_{\mathrm{s}} \dFM}.
\end{equation}

\begin{figure*}[tpb]
\center
 \includegraphics[width= \linewidth]{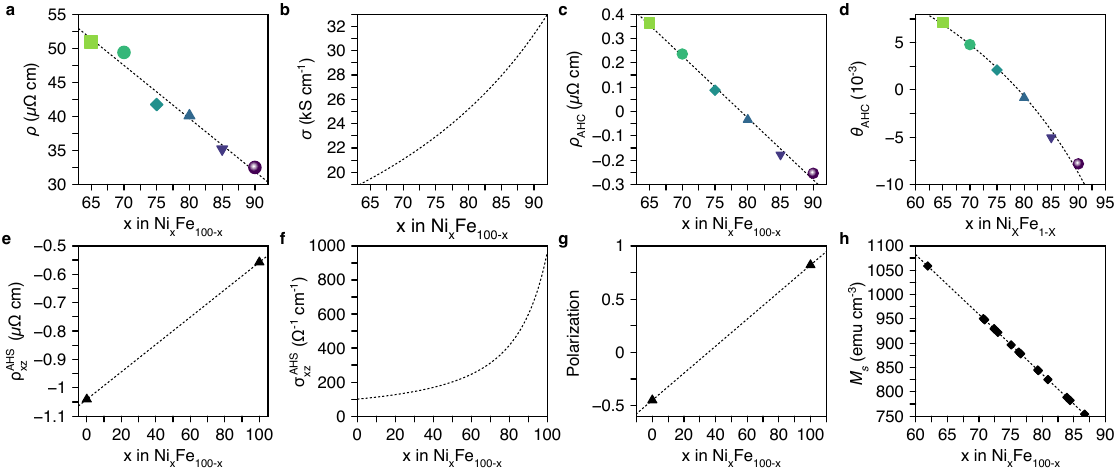}%
 \caption{\textbf{AHT damping efficiency $\Ayz$ as function of $\NiXFe$ alloy composition.} \textbf{(a)}  Longitudinal resistivity versus alloy composition $x$: experimental data (symbols) and linear fit. 
 \textbf{(b)} Longitudinal conductivity $\sigma$ versus $x$. 
 \textbf{(c)} Anomalous Hall resistivity versus alloy composition: experimental data (symbols) and linear fit. 
 \textbf{(d)} Anomalous Hall angle versus $x$ calculated from the data in (a) and (c).  
 \textbf{(e)} Linear interpolation of transverse spin resistivity (line) based on its values for Ni and Fe (triangles) calculated in Ref.\,\cite{aminIntrinsicSpin2019}. 
 \textbf{(f)}  Transverse spin conductivity calculated from the line in (e). \textbf{(g)} Linear interpolation of longitudinal current polarization $P$ (line) based on its values for Ni and Fe (triangles) calculated in Ref.\,\cite{kokadoAnisotropicMagnetoresistance2012}. \textbf{(h)} Linear fit (line) of saturation magnetization data (diamonds) for $\NiXFe$ films taken from Ref.\,\cite{boninDependenceMagnetization2005}.
\label{fig:Ayz_estimation}}%
 \end{figure*}

We now calculate the dependence of $\Ayz$ on the $\NiXFe$ alloy composition. We start by evaluating $\thetaAHT = \thetaX - P \thetaAHE$ as a function of $x$. First we estimate a smooth function for $\thetaAHE$. We fit the longitudinal and anomalous Hall charge resistivities, $\rho$ and $\rhoAHE$, with a linear function with the results shown as dotted black lines in \cref{fig:Ayz_estimation} (a) and (c) respectively. \Cref{fig:Ayz_estimation}(b) shows the longitudinal conductivity $ \sigma\equiv \sigmaxx$ calculated using the inverse of the fit function $\rho (x)$: $\sigma (x) = \rho^{-1}(x)$. \Cref{fig:Ayz_estimation}(d) shows $\thetaAHE(x) = \rhoAHE(x) / \rho(x)$ as the black dotted line.

Next, we estimate the polarization $P$ for Ni and Fe based on the majority ($\rho_\uparrow$) and minority ($\rho_\downarrow$) channel resistivities found in Ref. \cite{kokadoAnisotropicMagnetoresistance2012}: $\frac{(\rho_\downarrow)}{(\rho_\uparrow)} = 10$ for Ni and $\frac{(\rho_\downarrow)}{(\rho_\uparrow)} = 0.38$ for Fe. Using the relation,
\begin{equation}
    P=\frac{\rho_\downarrow - \rho_\uparrow }{\rho_\downarrow + \rho_\uparrow } = \frac{\frac{\rho_\downarrow}{\rho_\uparrow} - 1}{\frac{\rho_\downarrow}{\rho_\uparrow} + 1},
    \label{eq:polar}
\end{equation}
we estimate the $P_\mathrm{Ni} = 0.82$ and  $P_\mathrm{Fe} = -0.45$. These values imply that the spin current in Ni is mainly carried by majority s-band electrons while in Fe the spin current is mainly carried by minority s-band electrons. We linearly interpolate between $P_\mathrm{Ni}$ and $P_\mathrm{Fe}$ to estimate the polarization for the alloys; the resulting $P(x)$ is shown as the black dotted line in \cref{fig:Ayz_estimation}(g). We plot the term $-P(x)\thetaAHE(x)$ as the red curve in Fig. 4(c) of the main text. 

Next we estimate the dimensionless parameter $\thetaX$. There are no experiments reporting values of $\thetaX$ or $\sigmaxzX$; however, \citet{aminIntrinsicSpin2019} used density functional theory to calculate $\sigmaxzX$ for Ni and Fe. 
Using the notation from Ref.\,\cite{aminIntrinsicSpin2019} on the right-hand-side, $\sigmaxzX = \sigmaSHE + \sigmaSAHE$. 
The calculated values reported in \cite{aminIntrinsicSpin2019} are: $\sigmaSHE^\mathrm{Ni} = 1688 \, \mathrm{\Omega^{-1} cm^{-1}}$, $\sigmaSAHE^\mathrm{Ni} = -728 \, \mathrm{\Omega^{-1} cm^{-1}}$, $\sigmaSHE^\mathrm{Fe} = 519 \, \mathrm{\Omega^{-1} cm^{-1}}$, and $\sigmaSAHE^\mathrm{Fe} = -419 \, \mathrm{\Omega^{-1} cm^{-1}}$. 
Therefore we estimate $\sigmaxzXNi = 960 \, \mathrm{\Omega^{-1} cm^{-1}} $ and $\sigmaxzXFe = 100 \, \mathrm{\Omega^{-1} cm^{-1}} $. 
We note that it is the resistivities of the experimentally measured parameters that have shown a linear dependence on Ni alloy concentration $x$. 
Therefore we convert the calculated $\sigmaxzX$ values into corresponding resistivities $\rhoxzX$ using the relation: $\rhoxzX = -\sigmaxzX \rho^2$. 
We find $\rhoxzXNi = -0.558 \, \mathrm{\mu \Omega \, cm }$ and $\rhoxzXFe = -1.04\, \mathrm{\mu \Omega \, cm }$. 
We linearly interpolate between these values of to find $\rhoxzX(x)$ as shown in \cref{fig:Ayz_estimation}(e). 
The corresponding $\sigmaxzX(x) = -\rhoxzX(x) \rho(x)^2$ is shown in \cref{fig:Ayz_estimation}(f). Using \cref{eq:S38,eq:S39} we evaluate the term $\thetaX(x) = \sigmaxzX(x) / \sigma(x)$. 
The resulting $\thetaX(x)$ is shown as the blue curve in Fig.\,4(c) of the main text. 
The net dimensionless anomalous Hall torque efficiency $\thetaAHT(x) = \thetaX(x) - P \thetaAHE(x)$ is shown as the purple curve in Fig.\,4(c) of the main text. 
Qualitatively, the magnitude of $\thetaAHT$ increases with increasing Ni concentration in good agreement with increasing torque strength found in our experiments.

As $\Ayz$ is inversely proportional to $\Ms$, we numerically estimate the alloy dependence of $\Ms$ via a linear fit of the data presented in Table I of Ref. \cite{boninDependenceMagnetization2005}. 
The data and resulting fit are shown as the black diamonds and black dotted line in \cref{fig:Ayz_estimation} (h). The resulting dependence of $\Ayz$ on NiFe alloy concentration,
\begin{equation}
    \Ayz(x) = \frac{\hbar}{8 |e|}\frac{\thetaAHT(x)}{\Ms (x) \dFM} ,
\end{equation}
is shown in Fig.\,2c of the main text, where $\dFM = 5 \, \mathrm{nm}$.

\section{Landau Lifshitz Gilbert equation with anomalous Hall torque and planar Hall torque terms}\label{sec:LLGextended}

The general method for predicting the impact of SHT, AHT and PHT on magnetization dynamics in FM$|$NM bilayers \cite{letang_modulation_2019, grollierNeuromorphicSpintronics2020, flebus_non-hermitian_2020, juge_helium_2021} is solving the LLG equation:
\begin{equation}\label{eq:LLGextended}
\frac{\partial \mathbf{M}}{\partial t}=-\gamma[\mathbf{M} \times \mathbf{H}_\mathrm{eff}] +\frac{\alpha}{M_\mathrm{s}}[\mathbf{M} \times \frac{\partial \mathbf{M}}{\partial t}] +\mathbf{T}_\mathrm{SHT} +\mathbf{T}_\mathrm{AHT} +\mathbf{T}_\mathrm{PHT},
\end{equation}
where $\gamma$ is the gyromagnetic ratio and $\mathbf{H}_\mathrm{eff}$ is the effective magnetic field acting on the FM layer. 
The last three terms in \cref{eq:LLGextended} are SHT, AHT and PHT. 
Below we give explicit expressions for $\mathbf{T}_\mathrm{SHT}$, $\mathbf{T}_\mathrm{AHT}$ and $\mathbf{T}_\mathrm{PHT}$ \cite{ochoaSelfinducedSpinorbit2021a}:
\begin{equation}\label{eq:TorqeeFormSHT}
    \mathbf{T}_{\mathrm{SHT}} = \gamma\frac{\hbar}{2 |e| }\frac{\vartheta_\mathrm{SHE} J_{\mathrm{NM}}}{\dFM }\left[ \mathbf{m} \times \left[  \mathbf{m} \times \hat{\mathbf{y}} \right] \right],
\end{equation}

\begin{equation}\label{eq:TorqeeFormAHT}
\mathbf{T}_\mathrm{AHT}=  -\gamma \frac{\hbar}{2 |e|} \frac{\thetaAHT}{d_{FM}} m_z (\bf{m} \times \hat{z} \times \bf{m})(\bf{m} \cdot (\hat{z} \times \bf{\JFM})),
\end{equation}

\begin{equation}\label{eq:TorqeeFormPHT}
\mathbf{T}_\mathrm{PHT}=  -\gamma \frac{\hbar}{2 |e|} \frac{ \thetaPHT}{d_{FM}} m_z^2 (\bf{m} \times \hat{z} \times \bf{m}) (\bf{m} \cdot  \bf{\JFM}).
\end{equation}

In \cref{eq:TorqeeFormPHT}, the dimensionless PHT efficiency parameter $\thetaPHT$ is given by \cite{ochoaSelfinducedSpinorbit2021a}:
\begin{equation}\label{eq:PHTEff}
\thetaPHT = \rho_\mathrm{AMRS}-P\rho_\mathrm{AMR},
\end{equation}
where $\rho_\mathrm{AMR}$ is the AMR ratio of the FM, $P$ is given by \cref{eq:polar} and $\rho_\mathrm{AMRS}$ is the spin sector analog of $\rho_\mathrm{AMR}$:  $\rho_\mathrm{AMRS}$ describes the transverse spin current driven by the same SOC that gives rise to the planar Hall charge current and AMR \cite{ochoaSelfinducedSpinorbit2021a}. 
We note that \cref{eq:TorqeeFormAHT} and \cref{eq:TorqeeFormPHT} are defined for bilayers where the good spin sink is on the FM film bottom and poor spin sink is on film top, as the case for our devices. For good spin sink on top and poor spin sink on the bottom of the FM film, the sign of \cref{eq:TorqeeFormAHT} and  \cref{eq:TorqeeFormPHT} must be changed. Similarly \cref{eq:TorqeeFormSHT} is defined for bilayer where the spin Hall material is on the FM film bottom. For spin Hall material on top the FM film, the sign of \cref{eq:TorqeeFormSHT} must be changed. Additionally, we note that \cref{eq:TorqeeFormAHT} is more precise than the main text Eq.\,(3) because the effect of magnetic anisotropy on the FMR linewidth is neglected in Eq.\,(3) while this approximation is not used in  \cref{eq:TorqeeFormAHT} \cite{ochoaSelfinducedSpinorbit2021a}.

For some FM$|$NM systems, additional SOT terms should be added to \cref{eq:LLGextended} \cite{alghamdi_highly_2019, bonell_control_2020, finizio_time-resolved_2020, chen_connections_2022}, including self-generated SOTs \cite{kurebayashiAntidampingSpin2014, jiangEfficientFull2019, luoSpinOrbitTorque2019, wangAnomalousSpin2019, satoEvaluationSpin2019, leeSpinOrbitTorque2020, tangBulkSpin2020, sekiSpinorbitTorque2021, chenCurrentinducedMagnetization2021, cespedes-berrocalCurrentInducedSpin2021, dongCurrentInducedMagnetic2022a, leeEffectsSelftorque2022, liEfficientMagnetization2023,aokiGiganticAnisotropy2023,martin_strong_2023}. 
In contrast, to self-generated AHT and PHT, such terms are generally non-universal and depend on a particular crystal symmetry or a chemical composition gradient.

\section{Comparison of measured and calculated critical currents for anomalous Hall torque nano-oscillator}\label{sec:Jc0}

Here we numerically evaluate the critical current density for the AHT nano-oscillator in Fig.\,3 of the main text (Ta(5\,nm)$|\NiFe{90}{10}$(5\,nm)$|$AlO$_\mathrm{y}$(2\,nm)) using \cref{eq:AHT_critital_current}. 
We take the values of $\alpha = 0.0149$ and $g = 2.18$ from our film-level FMR measurements as shown in \Cref{table2} and use $\dFM=5 \,\mathrm{nm}$. 
We use $M_\mathrm{s} = 714 \, \mathrm{emu \,cm}^{-3}$ appropriate for \NiFe{90}{10} \cite{boninDependenceMagnetization2005}. 
The anomalous Hall torque oscillator shown in Fig.\,3 of the main text emits microwave radiation near $\omega = 2 \pi \, 3.5 \, \mathrm{GHz}$ at the critical current. 
The magnetic field was applied at $\theta = 35^\circ$ and $\phi = 90^\circ$. We estimate $\thetaAHT = 0.025$ as described in Supplementary Note 7 and plotted in Fig. 4(c) of the main text. 
Substituting these values into \cref{eq:AHT_critital_current}, we estimate the theoretically expected value of $J_{\mathrm{FM0}}$:
\begin{equation}
    J_{\mathrm{FM0}}^{\mathrm{th}}  = 1.95 \times 10^{12} \mathrm{A \, m^{-2}}.
\end{equation}
The critical current of the anomalous Hall oscillator shown in Fig.\,3 of the main text is approximately $I_\mathrm{dc}^0=0.75$\,mA. 
Using parallel resistors model for data given in Table S3, we estimate that 87\% of the current flows through the $\NiFe{90}{10}$ layer.
The nanowire width is $w=56$\,nm and the FM thickness is $\dFM=5$\,nm. 
The experimental critical current density is therefore the current flowing through the $\NiFe{90}{10}$ layer at the onset of auto-oscillations divided by its cross-sectional area:

\begin{equation}
    J_{\mathrm{FM0}}^{\mathrm{exp}} = \frac{0.87\, I_\mathrm{dc}^0}{ w\,\dFM} = 2.3 \times 10^{12} \, \mathrm{A}\, \mathrm{m}^{-2}.
\end{equation}

The expected and the measured values of the AHT STNO critical currents are in good agreement. Novel materials with giant anomalous Hall currents \cite{shen33Giant2020, yangGiantUnconventional2020} and large $\thetaAHT$ can potentially greatly reduce the critical current density for AHT STNOs.

\bibliography{refer}